\documentclass[journal]{IEEEtran}

\usepackage{cite}
\ifCLASSINFOpdf
  \usepackage[pdftex]{graphicx}
\else
\fi
\usepackage{amsmath}
\usepackage{amsfonts}
\usepackage{bm}
\usepackage{dsfont}
\usepackage{algorithmic}
\ifCLASSOPTIONcompsoc
 \usepackage[caption=false,font=normalsize,labelfont=sf,textfont=sf]{subfig}
\else
 \usepackage[caption=false,font=footnotesize]{subfig}
\fi
\usepackage[hidelinks]{hyperref}
\usepackage{graphicx}
\usepackage{tabularx}
\usepackage{algorithm}
\usepackage{amssymb}
\usepackage{breqn}
\usepackage{tikz}
\usetikzlibrary{bayesnet}
\usepackage{svg}
\usepackage{needspace}
\usepackage{epstopdf}
\newcommand{\Gaussian}{\mathcal{N}}
\newcommand{\aux}{\mathcal{L}}
\newcommand{\vect}[1]{{#1}}
\newcommand{\mathintext}[1]{$#1$}
\newcommand{\timeind}{k}

\newcommand{\targetind}{m}
\newcommand{\Targetind}{M_\timeind}
\newcommand{\pixelindx}{i}
\newcommand{\Pixelindx}{I}

\newcommand{\pixelind}{i}
\newcommand{\Pixelind}{I}
\newcommand{\itind}{p}

\newcommand{\transp}{\mathrm{\textsf{T}}}
\newcommand{\makesubandsuper}[3]{#1_{#2}^{#3}}
 
\newcommand{\blue}[1]{\textcolor{black}{#1}}

\begin{document}

\title{Histogram-Probabilistic Multi-Hypothesis Tracking with Integrated Target Existence}
\author{Lukas~Herrmann,
        Ángel~F.~García-Fernández,
        and Edmund~F.~Brekke
\thanks{This work was supported by the Norwegian Research Council under SFI AutoShip (project number 309230), and the European Union’s Horizon 2020 research and innovation program under the PERSEUS doctoral program (Marie Skłodowska-Curie grant agreement number 101034240).}
\thanks{Lukas Herrmann is with the Department of Electronic Systems, Norwegian
University of Science and Technology, Trondheim, Norway (email: lukas.herrmann@ntnu.no).}
\thanks{Ángel F. Garc\'ia-Fern\'andez is with ETSI de Telecomunicaci\'on, Universidad Polit\'ecnica de Madrid, 28040 Madrid, Spain (email: \mbox{angel.garcia.fernandez@upm.es}).}
\thanks{Edmund F. Brekke is with the Department of Engineering Cybernetics, Norwegian
University of Science and Technology, Trondheim, Norway (email: edmund.brekke@ntnu.no).}}

\markboth{IEEE Transactions on XXX,~Vol.~XX, 2024}
{Shell \MakeLowercase{\textit{et al.}}: Bare Demo of IEEEtran.cls for IEEE Journals}

\maketitle

\begin{abstract}
The histogram-probabilistic multi-hypothesis track-er (H-PMHT) is a parametric approach to solving the multi-target track-before-detect (TBD) problem, using expectation maximisation (EM). A key limitation of this method is the assumption of a known and constant number of targets. In this paper, we propose the integrated existence Poisson histogram probabilistic multi-hypothesis tracker (IE-PHPMHT), for TBD of multiple targets. It extends the H-PMHT framework by adding a probability of existence to each potential target. For the derivation, we utilise a Poisson point process (PPP) measurement model and Bernoulli targets, allowing for a multi-Bernoulli birth process and an unknown, time-varying number of targets. Hence, integrated track management is achieved through the discrimination of track quality assessments based on existence probabilities. 
The algorithm is evaluated in a simulation study of two scenarios and is compared with several other algorithms, demonstrating its performance.
\end{abstract}


\IEEEpeerreviewmaketitle

\section{Introduction}
\IEEEPARstart{M}{ultiple} target tracking \cite{blackman_design_1999} has been an important field of research for decades as it is of great interest for a variety of applications such as autonomous vehicles, image processing, navigation, and maritime collision avoidance. A core challenge in multi-target tracking is the fact that the number of targets is both unknown and time-varying \cite{blackman_design_1999,bar-shalom_estimation_2001,mallick_integrated_2013}. Within the countless multi-target tracking methods, the most commonly used are based on multiple hypothesis tracking \cite{reid_algorithm_1979}, joint probabilistic data association \cite{fortmann_multi-target_1980}, particle filters \cite{hue_tracking_2002}, and \textit{random finite sets} (RFSs) \cite{mahler_statistical_2007}.

In many cases of interest, the available \textit{signal-to-noise ratio} (SNR) is so low that there cannot be found any reasonable threshold for the creation of point detections in the classical two-step procedure where the detection step is followed by subsequent tracking. This has led to the development of a variety of \textit{track-before-detect} (TBD) approaches (see e.g., \cite{salmond_particle_2001, rutten_recursive_2005, kreucher_multitarget_2005,morelande_bayesian_2007, buzzi_track-before-detect_2008, garcia-fernandez_track-before-detect_2016, papi_particle_2015}). In TBD, the full raw data is fed as an input to the tracking algorithm \cite{mallick_integrated_2013}. By avoiding the use of thresholds, all available information can be utilised, making TBD particularly interesting in low SNR scenarios, which then can lead to improved performance \cite{davey_comparison_2008}.

In multi-target tracking, a conventional approach to address uncertainty in the number of targets is to use an RFS formulation, where the multi-object state is represented as a set of targets \cite{mahler_advances_2014}. Of particular relevance to this paper is the use of multi-Bernoulli RFSs \cite{mahler_advances_2014,garcia-fernandez_gaussian_2019,ristic_tutorial_2013,vo_cardinality_2009}, where each element within the set is a Bernoulli component. Multi-Bernoulli TBD filters were proposed with both particle filter \cite{garcia-fernandez_track-before-detect_2016, papi_particle_2015, kim_bernoulli_2021} and Gaussian implementations\cite{davies_information_2024}. 

In \cite{streit_tracking_2000}, the \textit{histogram-probabilistic multi-hypothesis tracker} (H-PMHT) has been proposed as an TBD version of the \textit{probabilistic multi-hypothesis tracker} (PMHT) \cite{streit_probabilistic_1995}. Among multi-target TBD methods, H-PMHT has earned recognition as it is computationally very efficient \cite{davey_comparison_2008}, which is why it is the focus of this paper. In recent years, H-PMHT has been studied and extended \cite{davey_histogram-pmht_2013,wieneke_histogram-pmht_2014,davey_track-before-detect_2018}, and the latest extension of the H-PMHT, referred to as the \textit{Poisson} H-PMHT (PHPMHT), has re-derived the H-PMHT based on a Poisson measurement model and simultaneously proposed a way to handle target fluctuations \cite{gaetjens_histogram-pmht_2017}.

Although it has significant practical implications, a rarely analysed area of research has been the fact that both the H-PMHT and PHPMHT algorithms assume a fixed and known number of targets. The challenge of managing potential targets, including the initiation and termination of tracks, is commonly referred to as track management.
Most of the existing research in this domain has been focused on the non-TBD version of PMHT. In \cite{luginbuhl_track_2001}, the Radon transform was employed for track initiation, while track termination was handled through a threshold applied to the number of measurements assigned to each target. In \cite{davey_integrated_2007}, a method was proposed to discriminate between valid and false tracks using an integrated hysteresis approach \blue{where the track management quality score was termed target visibility. In this paper, we adopt the term existence as it is well-established in the multi-Bernoulli literature \cite{garcia-fernandez_trajectory_2020, vo_joint_2010, davies_information_2024,garcia-fernandez_gaussian_2019,ristic_tutorial_2013,vo_cardinality_2009, kim_bernoulli_2021}. If visibility is understood to describe factors such as occlusions, the existence model represents the physical existence while detectability is included in the probability of detection for detection-based models.}
Similarly, \cite{musicki_track_2007} developed a model that combined prior association probabilities with a hidden Markov model to enhance track management. 
\blue{In \cite{turner_complete_2014} a variational Bayes tracker for a point target detection model was proposed to address the integrated track management problem, enforcing that the posterior factorises across assignment matrices and target states.}
Another extension of PMHT, based on sequential likelihood ratio testing, was introduced in \cite{wieneke_sequential_2007} and subsequently extended with a Gaussian mixture probability hypothesis density filtering step to mitigate the effects of clutter, as discussed in \cite{wieneke_track-management_2008}. Within the H-PMHT framework, the issue of track management has only been briefly addressed in the available literature, typically as a secondary aspect of the algorithm \cite{davey_comparison_2008,davey_detecting_2011,mallick_integrated_2013}. The few practical extensions that exist share a common approach. They proposed to directly estimate the SNR from the H-PMHT mixture proportions for the mechanism of initiating and terminating tracks. Track initiation is typically based on peaks detected in a prior detection step, while track management depends on SNR estimates for each potential target. The SNR-estimates must exceed a confirmation threshold to promote a track or fall below a termination threshold to remove it. Additionally, an M/N logic was introduced to complement the SNR-based threshold methods, which otherwise lack temporal correlation. An additional dynamic programming based detection step combined with H-PMHT was proposed in \cite{zhang_h-pmht_2016} and in a more recent development, \cite{guo_sa-hpmht_2023} applied additional track management techniques using a constant false alarm detector, alongside SNR estimates and thresholding, to a model accounting for sensor uncertainty. For the PHPMHT framework, no track management extensions have been reported in the literature.

In this paper, we propose a new PHPMHT algorithm with integrated probability of existence. We utilise a PPP measurement model and incorporate target existence probabilities, leading to a new formulation, which we refer to as \textit{integrated existence} PHPMHT (IE-PHPMHT). The model extends the PHPMHT framework by integrating target existence probabilities, allowing it to manage an unknown and time-varying number of targets through built-in track management. The IE-PHPMHT introduces Bernoulli targets, and in order to maintain a closed-form solution and low computational complexity, we approximate the resulting distribution of the Poisson rate by minimising the \textit{Kullback-Leibler divergence} (KLD).
Furthermore, we provide the derivation by introducing auxiliary variables in the update step. Once the approximated joint density of the measurements, auxiliary variables, Poisson rates, and states are formulated, the most likely states are obtained via \textit{expectation maximisation} (EM) \cite{bishop_pattern_2006}. This forms a concise derivation of the IE-PHPMHT, including an alternative derivation of the relevant parts of the PHPMHT.

The rest of the paper is organised as follows. In Section \ref{sec:ProblemFormulation} we formulate the considered multi-target tracking problem. Section \ref{sec:ExpectationMaximisation} presents a solution to the problem by using EM. The resulting IE-PHPMHT with integrated target existence probabilities is derived in Section \ref{sec:IE-HPMHT}. A simulation study is carried out in Section \ref{sec:SimulationResults} and the results are analysed. Finally, conclusions are drawn in Section \ref{sec:Conclusion}.

\section{Problem Formulation}
\label{sec:ProblemFormulation}
In this section, we provide the problem formulation for multi-target tracking within the H-PMHT framework \cite{streit_tracking_2000,streit_multitarget_2002,luginbuhl_estimating_2004,gaetjens_histogram-pmht_2017} for an unknown and time-varying number of targets. In particular, we introduce an existence variable to each target state with its corresponding probability of existence. We present the model for the targets' dynamics in Section \ref{sec:Problem_dynamics}, the measurement model in Section \ref{sec:Problem_measurement}, the prediction in \ref{sec:Problem_pred} and the resulting joint distribution for the update step in Section \ref{sec:Problem_jointdist}. 

Assuming that, at each discrete time step \mathintext{k}, there are \mathintext{\Targetind} targets moving independently, we aim to estimate the multi-target state \mathintext{\makesubandsuper{X}{\timeind}{}~=~ [\makesubandsuper{\vect{x}}{\timeind}{1},...,\makesubandsuper{\vect{x}}{\timeind}{\targetind},...,\makesubandsuper{\vect{x}}{\timeind}{\Targetind} ]^{\transp}} which is given by the collection of the \mathintext{\mathcal{M}_\timeind~=~\{1, ..., \Targetind\}} state vectors \mathintext{\makesubandsuper{\vect{x}}{\timeind}{\targetind} \in \mathbb{R}^{d_x}}, where \mathintext{d_x} is the dimension of the states and  \mathintext{m \in \mathcal{M}_\timeind} is the target index.
We are given a vector of histogram data \mathintext{N_\timeind~=~\left[ \makesubandsuper{n}{\timeind}{1},..., \makesubandsuper{n}{\timeind}{\pixelind},...,\makesubandsuper{n}{\timeind}{\Pixelind} \right]^\transp} where every element contains a random variable representing the number of measurements \mathintext{\makesubandsuper{n}{\timeind}{\pixelind}} that are located inside each cell \mathintext{i}, and \mathintext{\Pixelind} is the total number of resolution cells.  
\subsection{Dynamic Models}
\label{sec:Problem_dynamics}
In order to model appearing and disappearing targets, the presence or absence of each target is modelled by a discrete binary random variable \mathintext{\makesubandsuper{e}{\timeind}{\targetind} \in \{0,1}\}. We follow the intuitive convention where \mathintext{\makesubandsuper{e}{\timeind}{\targetind} = 1} indicates that the \mathintext{m}-th target is present at time \mathintext{k}, and conversely, \mathintext{\makesubandsuper{e}{\timeind}{\targetind} = 0} indicates its absence.

We consider a multi-Bernoulli birth model in which \mathintext{M_\timeind^b} new targets can appear at each time step. The \mathintext{\targetind^b}-th new target is born with a probability of \textit{target birth} \mathintext{p_b} and a single-target birth density \mathintext{b(\cdot)} \cite{mahler_advances_2014},

\begin{align}
\label{eq:existence_prior_birth}
    p(\makesubandsuper{e}{\timeind}{\targetind^b}|\makesubandsuper{e}{\timeind-1}{\targetind^b}) = 
    \begin{cases}
        p_b \quad & \mathrm{if}\ \makesubandsuper{e}{\timeind}{\targetind^b} = 1,\ \makesubandsuper{e}{\timeind-1}{\targetind^b} = 0 \\
        1 - p_b \quad & \mathrm{if}\ \makesubandsuper{e}{\timeind}{\targetind^b} = 0,\ \makesubandsuper{e}{\timeind-1}{\targetind^b} = 0 \\
        0 \quad & \text{otherwise}
    \end{cases}\;.
\end{align}

The targets already present at the previous time step \mathintext{\timeind-1} are considered to move independently and can either continue to exist or disappear at the current time step \mathintext{k}. Thus, the dynamics of the existence of the surviving targets is modelled by a first-order Markov chain and the transition prior density is given by:
\begin{align}
\label{eq:existence_prior}
    p(\makesubandsuper{e}{\timeind}{\targetind}|\makesubandsuper{e}{\timeind-1}{\targetind}) = 
    \begin{cases}
        p_s \quad & \mathrm{if}\ \makesubandsuper{e}{\timeind}{\targetind} = 1,\ \makesubandsuper{e}{\timeind-1}{\targetind} = 1 \\
        1 - p_s \quad & \mathrm{if}\ \makesubandsuper{e}{\timeind}{\targetind} = 0,\ \makesubandsuper{e}{\timeind-1}{\targetind} = 1 \\
        0 \quad & \text{otherwise}
    \end{cases}\;,
\end{align}
where \mathintext{p_s} is the probability of \textit{target survival}. Given a target that survives from the last time step to the current, its state is distributed according to a known state transition density \mathintext{p(\makesubandsuper{\vect{x}}{\timeind}{\targetind}|\makesubandsuper{\vect{x}}{\timeind-1}{\targetind})}.

Both \mathintext{p_s} and \mathintext{p_b} are assumed to be stationary and equal for all targets and the total number of potential targets at time step \mathintext{k} is the sum of surviving targets and newborn targets i.e., \mathintext{\Targetind = M^s_{\blue{k-1}} + \Targetind^b}.

The collection of the existence variables of all \mathintext{\Targetind} potential targets is denoted \mathintext{E_\timeind~=~\left[ \makesubandsuper{e}{\timeind}{1},...,\makesubandsuper{e}{\timeind}{\targetind},...,\makesubandsuper{e}{\timeind}{\Targetind} \right]^{\transp}}.
\subsection{Measurement Model}
\label{sec:Problem_measurement}
\blue{The fundamental basis for H-PMHT is that the measurements are observations of an underlying mixture process with contributions from targets and clutter. For the derivation, the intensity in each resolution cell \mathintext{\pixelind} is considered quantised, resulting in the histogram data \mathintext{N_\timeind} which is treated as a realisation of a point process \cite{streit_tracking_2000}.}
Similarly to the PHPMHT \cite{gaetjens_histogram-pmht_2017}, we model the measurements as a realisation of a PPP \cite{streit_poisson_2010}. This forms the basis of our measurement model formulation where the PPP is modelled as a mixture process with the \blue{overall} intensity \blue{function of the entire data} given as:  
\begin{align}
\label{eq:meas_mod_intensity}
    \nu_\timeind(y|X_\timeind,\Lambda_\timeind) = \makesubandsuper{\lambda}{\timeind}{0} p^0(y) + \makesubandsuper{\sum}{\targetind=1}{\Targetind} \makesubandsuper{\lambda}{\timeind}{\targetind} p^\targetind(y|\makesubandsuper{\vect{x}}{\timeind}{\targetind})\;,
\end{align} where \mathintext{y} represents a measurement \blue{location} in the continuous measurement space \mathintext{\mathbb{R}^{n_y}}, \mathintext{\makesubandsuper{\lambda}{\timeind}{\targetind}} is the Poisson measurement rate of the \mathintext{m}-th component, the spatial distribution of the clutter measurements is given by the clutter density \mathintext{p^0(y)}, and the influence of each target on each measurement cell is modelled by the density \mathintext{\makesubandsuper{p}{}{\targetind}(y|\makesubandsuper{\vect{x}}{\timeind}{\targetind})}.
Note that with this notation the target index \mathintext{m=0} denotes clutter. The collection of the Poisson measurement rates of the mixture model is denoted as \mathintext{\Lambda_\timeind~=~\left[\makesubandsuper{\lambda}{\timeind}{0},..., \makesubandsuper{\lambda}{\timeind}{\targetind},...,\makesubandsuper{\lambda}{\timeind}{\Targetind}\right]^{\transp}}.

By integrating the intensity, the Poisson rate \mathintext{\blue{\makesubandsuper{\Bar{\nu}}{\timeind}{\pixelind}}} \blue{which depends on \mathintext{\blue{X_\timeind}} and \mathintext{\blue{\Lambda_\timeind}}} for \blue{the $i$-th} resolution cell can be found as:
\begin{align}
\label{eq:meas_mod_rate}
     \blue{\makesubandsuper{\Bar{\nu}}{\timeind}{\pixelind} \left( X_\timeind, \Lambda_\timeind \right)} = &\makesubandsuper{\lambda}{\timeind}{0} \int_{A^{\pixelind}} p^0(y) \mathrm{d}y + \makesubandsuper{\sum}{\targetind=1}{\Targetind} \makesubandsuper{\lambda}{\timeind}{\targetind} \int_{A^{\pixelind}} p^\targetind(y|\makesubandsuper{\vect{x}}{\timeind}{\targetind}) dy \;.
\end{align}
where \mathintext{A^{\pixelind}} indicates the spatial extend of resolution cell \mathintext{i}. Thus, the integrals in the above mixture are the probabilities of a measurement resulting from a certain component falling into a specific resolution cell. \blue{Note that \mathintext{\nu} in \eqref{eq:meas_mod_intensity} is a PPP intensity whereas \mathintext{\makesubandsuper{\Bar{\nu}}{\timeind}{\pixelind} \left( X_\timeind, \Lambda_\timeind \right)} is the Poisson rate of the mixture.}

For a given $X_\timeind$, $\Lambda_\timeind$, the Poisson rate of measurements in cell ${\pixelind}$ is given in (\ref{eq:meas_mod_rate}). 
This means that the \textit{probability mass function}~(PMF) of the number of measurements \blue{in the i-th resolution cell}, given $X_\timeind$  and \mathintext{\Lambda_\timeind}, is \blue{Poisson distributed}:
\begin{align}
p(n_\timeind^{{\pixelind}}) & =\text{e}^{-\blue{\Bar{\nu}_\timeind^{\pixelind}}}\frac{\left(\blue{\Bar{\nu}_\timeind^{\pixelind}}\right)^{n_\timeind^{\pixelind}}}{n_\timeind^{\pixelind}!}\label{eq:cardinality} \;,
\end{align}
where the explicit denoting of the conditioning has been removed for readability. It is worth mentioning that the number of measurements in each resolution cell follows a Poisson distribution as a result of the Poisson superposition property as the total number of measurements is Poisson distributed. Hence, without loss of generality, we also get, \mathintext{\makesubandsuper{\lambda}{\timeind}{\targetind} = \makesubandsuper{\sum}{\pixelindx=1}{\Pixelindx} \makesubandsuper{\lambda}{\timeind}{\pixelind,\targetind}} and \mathintext{||N_\timeind||\ = \makesubandsuper{\sum}{\pixelind=1}{\Pixelind}\makesubandsuper{n}{\timeind}{\pixelind} = n_\timeind} with \mathintext{||\cdot||} denoting the L1-norm and \mathintext{\makesubandsuper{n}{\timeind}{}} denoting the total number of measurements. \blue{The Poisson rate \mathintext{\makesubandsuper{\lambda}{\timeind}{\pixelind,\targetind}} represents the expected number of measurements in resolution cell \mathintext{\pixelind} originating from component \mathintext{\targetind}, directly linking the observed measurement count to the intensity of a target or clutter. The Poisson mixture model formulation is further used to address the data association problem.}

From (\ref{eq:meas_mod_rate}), we note that the observation is incomplete and we are concerned with two hidden variables, namely the component that gave rise to a measurement and the precise location of that measurement. Therefore, we use an auxiliary variable \cite{ubeda-medina_adaptive_2017,garcia-fernandez_trajectory_2020,bishop_pattern_2006} $m_\timeind\in\{0,1,...,\Targetind\}$ to indicate the index of the mixture component such that the \blue{intensity} of a PPP of variable \mathintext{m_\timeind} is:
\begin{align}
\blue{\nu_\timeind^{\pixelind} \left(m_\timeind|X_\timeind,\Lambda_\timeind\right)} 
=\begin{cases}
\lambda_\timeind^{0}\int_{A^{\pixelind}}p^{0}(y)dy & m_\timeind=0\\
\lambda_\timeind^{\targetind_\timeind}\int_{A^{\pixelind}}p^{\targetind_\timeind}(y|\vect{x}_\timeind^{\targetind_\timeind})dy & m_\timeind\in\mathcal{M}_\timeind \;.
\end{cases}
\end{align}
By integrating out $m_\timeind$ we recover the Poisson rate $\blue{\Bar{\nu}_\timeind^{\pixelind}}$, see (\ref{eq:meas_mod_rate}). Analogously, we can define another auxiliary variable $\makesubandsuper{\vect{y}}{\timeind}{\pixelind}\in A^{\pixelind}$ indicating the precise measurement location of one measurement \mathintext{\makesubandsuper{\vect{y}}{\timeind}{\pixelind}} such that the intensity of a PPP in the space of \mathintext{(m_\timeind, \vect{y}_\timeind^\pixelind)} is:
\begin{align}
\label{eq:measurement_model_wo_hidden}
&\blue{\nu_\timeind^{\pixelind}\left(m_\timeind,\makesubandsuper{\vect{y}}{\timeind}{\pixelind}|X_\timeind,\Lambda_\timeind\right)}  \nonumber\\ &= \begin{cases}
\lambda_\timeind^{0}p^{0}(\makesubandsuper{\vect{y}}{\timeind}{\pixelind}) \mathds{1}_{A^\pixelind}(\makesubandsuper{\vect{y}}{\timeind}{\pixelind}) & m_\timeind=0\\
\lambda_\timeind^{m_\timeind}p^{m_\timeind}(\makesubandsuper{\vect{y}}{\timeind}{\pixelind}|\vect{x}_\timeind^{m_\timeind}) \mathds{1}_{A^\pixelind}(\makesubandsuper{\vect{y}}{\timeind}{\pixelind}) & m_\timeind\in\mathcal{M}_\timeind \;,
\end{cases}
\end{align}
where \mathintext{\mathds{1}_{A^\pixelind}(\makesubandsuper{\vect{y}}{\timeind}{\pixelind})} is the indicator function which is one if the measurement \mathintext{\makesubandsuper{\vect{y}}{\timeind}{\pixelind}} is located inside the area \mathintext{A^\pixelind} of the resolution cell \mathintext{i} and zero otherwise. Again, integrating out both $m_\timeind$ and $\makesubandsuper{\vect{y}}{\timeind}{\pixelind}$ recovers the original Poisson rate, see (\ref{eq:meas_mod_rate}).

A sample from this PPP is a set of points $ \makesubandsuper{\mathcal{Y}}{\timeind}{\pixelind}~=~\left\{ \left(m_{\timeind}^{\pixelind,1},\vect{y}_{\timeind}^{\pixelind,1}\right),...,\left(m_{\timeind}^{\pixelind,n_\timeind^{\pixelind}},\vect{y}_{\timeind}^{\pixelind,n_\timeind^{\pixelind}}\right)\right\}$ with a multi-object density \cite{mahler_advances_2014}:
\begin{align}
\label{eq:multi_obj_dens_missing}
p\left( \makesubandsuper{\mathcal{Y}}{\timeind}{\pixelind} |X_\timeind,\Lambda_\timeind \right) = \text{e}^{-\blue{\Bar{\nu}_\timeind^{\pixelind}}}\prod_{r=1}^{n_\timeind^{\pixelind}}\blue{\nu_\timeind^{\pixelind}\left(m_{\timeind}^{\pixelind,r},\vect{y}_{\timeind}^{\pixelind,r}\right)} \;.
\end{align}
It should be noted that the cardinality of this set follows the Poisson distribution in (\ref{eq:cardinality}). Given $n_\timeind^{\pixelind}$, the density of the hidden variables \mathintext{\mathcal{Y}_\timeind^\pixelind} is:
\begin{align}
\label{eq:p_miss_pixel}
p\left( \makesubandsuper{\mathcal{Y}}{\timeind}{\pixelind} |n_\timeind^{\pixelind},X_\timeind,\Lambda_\timeind \right) &=\frac{\text{e}^{-\blue{\Bar{\nu}_\timeind^{\pixelind}}}\prod_{r=1}^{n_\timeind^{\pixelind}}\blue{\nu_\timeind^{\pixelind}\left(m_{\timeind}^{\pixelind,r},\vect{y}_{\timeind}^{\pixelind,r}\right)}}{p(n_\timeind^{\pixelind})}\nonumber\\
 & =\frac{n_\timeind^{\pixelind}!\prod_{r=1}^{n_\timeind^{\pixelind}}\blue{\nu_\timeind^{\pixelind}\left(m_{\timeind}^{\pixelind,r},\vect{y}_{\timeind}^{\pixelind,r}\right)}}{\left(\blue{\Bar{\nu}_\timeind^{\pixelind}}\right)^{n_\timeind^{\pixelind}}} \;.
\end{align}
Notice that this density is integrated via a set integral, but considering a fixed cardinality \cite{mahler_advances_2014}.

If we consider the distribution considering all resolution cells, we note that the random variable is a sequence of sets, which we denote as \mathintext{\mathbb{Y}_\timeind = \left( \makesubandsuper{\mathcal{Y}}{\timeind}{1},\cdots, \makesubandsuper{\mathcal{Y}}{\timeind}{\Pixelind} \right)}, such that:
\begin{align}
\label{eq:p_miss_complete}
p\left(\mathbb{Y}_\timeind|N_\timeind, X_\timeind,\Lambda_\timeind \right) & =\prod_{i=1}^{I} p\left( \makesubandsuper{\mathcal{Y}}{\timeind}{\pixelind} \right) \;.
\end{align}
Substituting (\ref{eq:measurement_model_wo_hidden}) into (\ref{eq:p_miss_pixel}) and (\ref{eq:p_miss_complete}) leads to the final expression:
\begin{align}
\label{eq:lh_measurements}
    p(\mathbb{Y}_\timeind|N_\timeind,X_\timeind,\Lambda_\timeind) 
    &= \makesubandsuper{\prod}{\pixelindx=1}{\Pixelindx} \frac{n_\timeind^{\pixelind}!\prod_{r=1}^{n_\timeind^{\pixelind}} \makesubandsuper{\lambda}{\timeind}{\makesubandsuper{m}{\timeind}{\pixelind,r}} \makesubandsuper{p}{}{i,\makesubandsuper{m}{\timeind}{\pixelind,r}}\left( \makesubandsuper{\vect{y}}{\timeind}{\pixelind,r}|\makesubandsuper{\vect{x}}{\timeind}{\makesubandsuper{m}{\timeind}{\pixelind,r}} \right)}{\left(\blue{\Bar{\nu}_\timeind^{\pixelind}}\right)^{n_\timeind^{\pixelind}}}\;.
\end{align}
\blue{To simplify notation in \eqref{eq:lh_measurements}, we have introduced the function:}
\everymath{\color{black}}
\begin{align}
    \makesubandsuper{p}{}{\makesubandsuper{i,m}{\timeind}{\pixelind,r}}\left( \makesubandsuper{\vect{y}}{\timeind}{\pixelind,r}|\makesubandsuper{\vect{x}}{\timeind}{\makesubandsuper{m}{\timeind}{\pixelind,r}} \right) = \makesubandsuper{p}{}{\makesubandsuper{m}{\timeind}{\pixelind,r}}\left( \makesubandsuper{\vect{y}}{\timeind}{\pixelind,r}|\makesubandsuper{\vect{x}}{\timeind}{\makesubandsuper{m}{\timeind}{\pixelind,r}} \right) \,
    \mathds{1}_{A^\pixelind}\left(\makesubandsuper{\vect{y}}{\timeind}{\pixelind,r}\right) \;.
\end{align}
\everymath{\color{black}}
\subsection{Prediction}
\label{sec:Problem_pred}
In this paper, we are interested in developing a multi-Bernoulli filter \cite{mahler_statistical_2007,vo_cardinality_2009,vo_joint_2010}. That is, the predicted density at time step \mathintext{k} can be written as:
\begin{equation}
    p(X_\timeind,E_\timeind) = p(X_\timeind)p(E_\timeind) \;,
\end{equation}
with
\begin{equation}
\label{eq:lh_exist}
    p(E_\timeind) = \makesubandsuper{\prod}{\targetind=0}{\Targetind} p(\makesubandsuper{e}{\timeind}{\targetind})
    \;.
\end{equation}

The distribution of each existence variable is given as:
\begin{align}
\label{eq:prob_exist}
    p(\makesubandsuper{e}{\timeind}{\targetind};\makesubandsuper{r}{\timeind}{\targetind}) = 
    \begin{cases}
        \makesubandsuper{r}{\timeind}{\targetind} &\quad \mathrm{if} \ \makesubandsuper{e}{\timeind}{\targetind} = 1 \\
        1 - \makesubandsuper{r}{\timeind}{\targetind} &\quad \mathrm{if} \ \makesubandsuper{e}{\timeind}{\targetind} = 0 \\
        0 &\quad \text{otherwise}
    \end{cases}\;,
\end{align}
with \mathintext{\makesubandsuper{r}{\timeind}{\targetind}} being the \textit{probability of existence}.

The multi-target state density is:
\begin{equation}
\label{eq:lh_state}
    p(X_\timeind) = \makesubandsuper{\prod}{\targetind=1}{\Targetind} p(\makesubandsuper{\vect{x}}{\timeind}{\targetind})
    \;.
\end{equation}

The prediction steps for multi-Bernoulli filter are well-known and can be summarised as follows \cite{garcia-fernandez_gaussian_2019}:
\begin{equation}
\label{eq:state_pred}
    p(\makesubandsuper{\vect{x}}{\timeind}{\targetind}) = \int p(\makesubandsuper{\vect{x}}{\timeind}{\targetind}|\makesubandsuper{\vect{x}}{\timeind-1}{\targetind}) p(\makesubandsuper{\vect{x}}{\timeind-1}{\targetind}) \, d\makesubandsuper{\vect{x}}{\timeind-1}{\targetind} \;,
\end{equation}
where \mathintext{p(\makesubandsuper{\vect{x}}{\timeind}{\targetind}|\makesubandsuper{\vect{x}}{\timeind-1}{\targetind})} is the state transition density, and
\begin{equation}
\label{eq:exist_pred}
        p(\makesubandsuper{e}{\timeind}{\targetind}) = \int p(\makesubandsuper{e}{\timeind}{\targetind}|\makesubandsuper{e}{\timeind-1}{\targetind}) p(\makesubandsuper{e}{\timeind-1}{\targetind}) \, d\makesubandsuper{e}{\timeind-1}{\targetind} \;.
\end{equation}
In case of a new born target, the predicted density is given by (\ref{eq:existence_prior_birth}). For surviving targets, the transition density \mathintext{p(\makesubandsuper{e}{\timeind}{\targetind}|\makesubandsuper{e}{\timeind-1}{\targetind})} for the prediction step is given by (\ref{eq:existence_prior}). 
\subsection{Joint Distribution for the Update}
\label{sec:Problem_jointdist}
In this paper, we focus on the update step. Hence, we proceed writing the joint density for all the variables involved in the update step. 

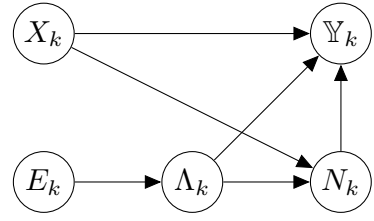
\begin{figure}[tb]
    \centering
    \begin{tikzpicture}[scale=1.15, transform shape] 
    \node[latent] (E) {$E_\timeind$};
    \node[latent, right=of E] (L) {$\Lambda_\timeind$};
    \node[latent, above=of E] (X) {$X_\timeind$};
    \node[latent, right=of L] (N) {$N_\timeind$};
    \node[latent, above=of N] (Y) {$\mathbb{Y}_\timeind$};
    \edge {E} {L};
    \edge {X} {N};
    \edge {X} {Y};
    \edge {L} {N};
    \edge {L} {Y};
    \edge {N} {Y};
    \end{tikzpicture}
    \caption{Bayesian network of the joint density for the proposed generative model of the measurements given the target states and their existences.}
    \label{fig:Bayes_net_model_2}
\end{figure}

In the context of jointly estimating the number of targets and their corresponding states, we are concerned with the joint density of the target states, the existences, the Poisson measurement rates, the number of measurements per cell, and the collection of sets containing the component that gave rise to each measurement as well as the precise measurement location. The Bayesian network in Fig. \ref{fig:Bayes_net_model_2} illustrates the generative model with the independence assumptions made for the joint distribution of the proposed method. Under these assumptions, the joint density and its factorisation is given as:
\begin{align}
\label{eq:p_comp_existence_1}
    &p(N_\timeind,\mathbb{Y}_\timeind,X_\timeind,\Lambda_\timeind,E_\timeind) \nonumber \\ 
    &= p(X_\timeind)p(E_\timeind)p(\Lambda_\timeind|E_\timeind)p(N_\timeind|X_\timeind,\Lambda_\timeind)p(\mathbb{Y}_\timeind|N_\timeind,X_\timeind,\Lambda_\timeind) \;. 
\end{align}

Note that one fundamental model assumption is reflected by the relationship between the target existence and the Poisson measurement rate. Essentially, the prior of the Poisson rate is dependent on the prior of the target existence. In other words, the more likely it is that a target exists the higher the chances that it produces a greater number of measurements and vice versa.   

The first and the second term in (\ref{eq:p_comp_existence_1}) are given by the output of the prediction step in (\ref{eq:lh_exist}) and (\ref{eq:lh_state}), respectively, whereas the third term is the conditional prior for the Poisson rates which are modelled as illustrated in Fig. \ref{fig:Bayes_net_model_2} and are given as:
\begin{equation}
\label{eq:lh_Prate}
    p(\Lambda_\timeind|E_\timeind) = \makesubandsuper{\prod}{\targetind=0}{\Targetind} p(\makesubandsuper{\lambda}{\timeind}{\targetind}|\makesubandsuper{e}{\timeind}{\targetind}) \;.
\end{equation}

The fourth term in (\ref{eq:p_comp_existence_1}) follows from (\ref{eq:cardinality}) and the product over all the resolution cells due to the PPP measurement model which implies independence of the number of measurements that fall in disjoint cells:
\begin{align}
\label{eq:lh_assignments_complete}
    p(N_\timeind|X_\timeind,\Lambda_\timeind) = \makesubandsuper{\prod}{\pixelindx=1}{\Pixelindx} \text{e}^{-\blue{\Bar{\nu}_\timeind^{\pixelind}}}\frac{\left(\blue{\Bar{\nu}_\timeind^{\pixelind}}\right)^{n_\timeind^{\pixelind}}}{n_\timeind^{\pixelind}!} \;.
\end{align}
Lastly, the final term in (\ref{eq:p_comp_existence_1}) is given in (\ref{eq:lh_measurements}).

We observe that, in order to estimate the states and the Poisson measurement rates, we can simplify the joint density by marginalising out the existence such that it can be written as:
\begin{align}
\label{eq:p_comp_existence_2}
    p&(N_\timeind,\mathbb{Y}_\timeind,X_\timeind,\Lambda_\timeind) \nonumber \\
    &= \sum_{E_\timeind} p(N_\timeind,\mathbb{Y}_\timeind,X_\timeind,\Lambda_\timeind,E_\timeind) \\ 
    \label{eq:p_comp_existence_2_expand}
    &=p(X_\timeind)p(\Lambda_\timeind)p(N_\timeind|X_\timeind,\Lambda_\timeind)p(\mathbb{Y}_\timeind|N_\timeind,X_\timeind,\Lambda_\timeind)
    \;.
\end{align}

The calculation of the marginalisation in (\ref{eq:p_comp_existence_2}), where \mathintext{p(\Lambda_\timeind) = \sum_{E_\timeind} p(\Lambda_\timeind|E_\timeind) p(E_\timeind)}, for obtaining a closed-form solution will be discussed in Section \ref{sec:HPMHT_KLDmin}. 

\section{Expectation Maximisation for MAP Estimation}
\label{sec:ExpectationMaximisation}
In this section, we formulate the general solution of how EM is used to obtain \textit{maximum a-posteriori} (MAP) estimates to the presented multi-target tracking problem. Particularly, besides the state and Poisson rate estimation in Section \ref{sec:EM_state_Poisson}, in Section \ref{sec:HPMHT_KLDmin} we derive an optimal approximation of the prior for the Poisson measurement rates via KLD minimisation to retain a closed-form structure in the EM update under the proposed existence-based formulation.
\subsection{State and Poisson Rate Estimation}
\label{sec:EM_state_Poisson}
We consider the previously formulated problem where we observe a collection of integer-valued numbers \mathintext{N_\timeind}, each element representing the number of measurements that fall into a certain resolution cell within the sensor's field of view. The goal is to obtain the MAP estimates of the targets' states \mathintext{X_\timeind} and their Poisson rates \mathintext{\Lambda_\timeind} which are given by the mode of the posterior \mathintext{p(X_\timeind,\Lambda_\timeind|N_\timeind)}.
Thus, the estimation problem can be written as:
\begin{align}
    \left(\hat{X}_\timeind,\hat{\Lambda}_\timeind\right)_{\text{MAP}}
    &=\underset{X_\timeind,\Lambda_\timeind}{\text{argmax}} \, p(X_\timeind,\Lambda_\timeind|N_\timeind) \nonumber\\
    &=\underset{X_\timeind,\Lambda_\timeind}{\text{argmax}} \, p(X_\timeind,\Lambda_\timeind,N_\timeind) \nonumber\\
    &= \underset{X_\timeind,\Lambda_\timeind}{\text{argmax}} \, \int p_{\text{comp}}\left(X_\timeind,\Lambda_\timeind,N_\timeind,\mathbb{Y}_\timeind \right) \, \delta \mathbb{Y}_\timeind
    \;,
\end{align}
where the joint density in the last line is referred to as the complete data likelihood. Note that the integral here is a sequence of set integrals \cite{mahler_advances_2014} over the sequence of sets \mathintext{\mathbb{Y}_\timeind} as a result of our measurement model formulation.

To calculate the MAP estimates via EM\cite{bishop_pattern_2006}, a cost function, namely the \textit{auxiliary function} \mathintext{\mathcal{L}}, is formed where the missing data is already marginalised out of the complete data likelihood by using the previous estimate for the random variables to be estimated. This conditional expectation of the logarithm of the complete data likelihood over the missing data likelihood is then maximised. The obtained estimates are then used again forming the new auxiliary function. This process is repeated until a convergence criterion is reached.
\begin{align}
    \left(\hat{X}_\timeind^{\itind},\hat{\Lambda}_\timeind^{\itind}\right)_{\text{MAP}} &= \underset{X_\timeind,\Lambda_\timeind}{\text{argmax}} \ \makesubandsuper{\aux}{\timeind}{\itind}\left(X_\timeind,\Lambda_\timeind|\hat{X}_\timeind^{\itind-1},\hat{\Lambda}_\timeind^{\itind-1} \right)\;,
\end{align}
where the superscript \mathintext{\itind} explicitly denotes the iteration index of the EM method and the EM auxiliary function is given as:

\begin{align}
\label{eq:aux}
    \makesubandsuper{\aux}{\timeind}{\itind}&\left(X_\timeind,\Lambda_\timeind|\hat{X}_\timeind^{\itind-1},\hat{\Lambda}_\timeind^{\itind-1}\right) \nonumber\\
    =& \ \mathbb{E}_{\mathbb{Y}_\timeind|N_\timeind,\hat{X}_\timeind^{\itind-1},\hat{\Lambda}_\timeind^{\itind-1}}\left[\log \{ p_{\text{comp}}\left( N_\timeind, \mathbb{Y}_\timeind, X_\timeind, \Lambda_\timeind \right) \}\right] \nonumber \\
    =& \int \log \{ p_{\text{comp}}\left( N_\timeind, \mathbb{Y}_\timeind, X_\timeind, \Lambda_\timeind \right) \} \nonumber \\
    &\times p_{\text{miss}}(\mathbb{Y}_\timeind | N_\timeind, \hat{X}_\timeind^{\itind-1}, \hat{\Lambda}_\timeind^{\itind-1}) \, \delta \mathbb{Y}_\timeind\;.
\end{align}
The factorisation of the complete data likelihood is given in (\ref{eq:p_comp_existence_2_expand}) and the expression for the missing data likelihood has been derived in Section \ref{sec:Problem_measurement} through introducing the auxiliary variables to the measurement model and is given by (\ref{eq:p_miss_pixel}) and (\ref{eq:p_miss_complete}), respectively.
\subsection{Closed-form EM Update via KLD Minimisation}
\label{sec:HPMHT_KLDmin}
As shown in Section \ref{sec:Problem_jointdist}, in the joint likelihood used for applying EM, the target existence has already been marginalised out. This means that before using EM to obtain the estimates, the current marginal joint distribution has to be calculated at every time step. Therefore, the prior of the Poisson measurement rates in (\ref{eq:p_comp_existence_2}) is given by:
\begin{equation}
\label{eq:prior_lambda}
    p(\Lambda_\timeind) = \makesubandsuper{\prod}{\targetind=0}{\Targetind}p(\makesubandsuper{\lambda}{\timeind}{\targetind}) = \makesubandsuper{\prod}{\targetind=0}{\Targetind} \sum_{\makesubandsuper{e}{\timeind}{\targetind}} p(\makesubandsuper{\lambda}{\timeind}{\targetind}|\makesubandsuper{e}{\timeind}{\targetind})p(\makesubandsuper{e}{\timeind}{\targetind}) \;,
\end{equation}
where the prior of the existence is given in (\ref{eq:lh_exist}). However, it remains to choose appropriate distributions for the conditional density of the Poisson rates \mathintext{p(\makesubandsuper{\lambda}{\timeind}{\targetind}|\makesubandsuper{e}{\timeind}{\targetind})}. One intuitive way would be to define that if a target exists it contributes more measurements depending on its signal strength hence causing a Poisson rate greater than zero. If the target does not exist it should not produce any measurements. However, from a practical point of view, this way of modelling might not be appropriate for two reasons. First, this would mean that the true measurement rates are readily available and known which is not the case as we estimate them. Second, due to the data association uncertainty, it is practically inaccurate to declare a potential target no longer a target only if zero measurements have been assigned to that component. Therefore, we propose to choose the conditional density of the Poisson rate as either a gamma distribution: 
\begin{align}
\label{eq:p_cond_l_e1}
    p(\makesubandsuper{\lambda}{\timeind}{\targetind}|\makesubandsuper{e}{\timeind}{\targetind}=1) &= \mathcal{G}(\makesubandsuper{\lambda}{\timeind}{\targetind};\makesubandsuper{\alpha}{\timeind}{\targetind},\makesubandsuper{\beta}{\timeind}{\targetind}) \nonumber \\
    &= \frac{\left(\makesubandsuper{\beta}{\timeind}{\targetind}\right)^{\makesubandsuper{\alpha}{\timeind}{\targetind}}}{\Gamma(\makesubandsuper{\alpha}{\timeind}{\targetind})}\left(\makesubandsuper{\lambda}{\timeind}{\targetind}\right)^{\makesubandsuper{\alpha}{\timeind}{\targetind}-1} \text{e}^{-\makesubandsuper{\beta}{\timeind}{\targetind}\makesubandsuper{\lambda}{\timeind}{\targetind}} \;, 
\end{align}
in case the target exists or as an exponential distribution:
\begin{equation}
\label{eq:p_cond_l_e0}
    p(\makesubandsuper{\lambda}{\timeind}{\targetind}|\makesubandsuper{e}{\timeind}{\targetind}=0) = \mathcal{E}(\makesubandsuper{\lambda}{\timeind}{\targetind};\makesubandsuper{\gamma}{\timeind}{\targetind}) = \makesubandsuper{\gamma}{\timeind}{\targetind} \text{e}^{-\makesubandsuper{\gamma}{\timeind}{\targetind}\makesubandsuper{\lambda}{\timeind}{\targetind}} \;,
\end{equation}
in case the target does not exist. Here, \mathintext{\alpha} and \mathintext{\beta} are the shape and rate parameters of the gamma distribution, \mathintext{\gamma} is the rate parameter of the exponential distribution, and \mathintext{\Gamma(\cdot)} denotes the Gamma function. \blue{A strategy for initialising the gamma shape and rate parameters is discussed in \cite{gaetjens_histogram-pmht_2017,davey_track-before-detect_2018}. The rate parameter of $\blue{\mathcal{E}}$ should be selected based on the expected clutter rate, with a lower value of $\blue{\gamma}$ corresponding to stronger clutter. In contrast, we could observe more stable results avoiding an excessively large $\blue{\gamma}$, as it causes the exponential distribution to concentrate closely around zero.}

This modelling approach accounts for the error and uncertainty in the estimates caused by measurements that have been incorrectly assigned to a component. Conversely, when a target does not exist, we assume the Poisson rate to follow an exponential distribution. The rationale is that a lower measurement rate increases the probability that measurements have been falsely attributed to a potential target when the true source was clutter.

By using the conditional densities for the Poisson rates (\ref{eq:p_cond_l_e1}) and (\ref{eq:p_cond_l_e0}) together with (\ref{eq:prior_lambda}) we obtain:
\begin{align}
\label{eq:prior_lambda_ex}
    p(\makesubandsuper{\lambda}{\timeind}{\targetind}) =& 
    \makesubandsuper{r}{\timeind}{\targetind}\frac{\left(\makesubandsuper{\beta}{\timeind}{\targetind}\right)^{\makesubandsuper{\alpha}{\timeind}{\targetind}}}{\Gamma(\makesubandsuper{\alpha}{\timeind}{\targetind})}\left(\makesubandsuper{\lambda}{\timeind}{\targetind}\right)^{\makesubandsuper{\alpha}{\timeind}{\targetind}-1} \text{e}^{-\makesubandsuper{\beta}{\timeind}{\targetind}\makesubandsuper{\lambda}{\timeind}{\targetind}} \nonumber \\
    &+ (1-\makesubandsuper{r}{\timeind}{\targetind})\makesubandsuper{\gamma}{\timeind}{\targetind} \text{e}^{-\makesubandsuper{\gamma}{\timeind}{\targetind}\makesubandsuper{\lambda}{\timeind}{\targetind}} \;.
\end{align}
As we can see from the above equation, the prior of the Poisson rate is the sum of a gamma distribution and an exponential distribution weighted by the probability of existence and non-existence, respectively. In this paper, we therefore propose to fuse this mixture into a single gamma distribution to keep the closed-form structure of the EM update.

To find the optimal approximation, we use KLD minimisation \cite{bishop_pattern_2006}, and the mixture merging is done according to the following Lemma 1 which formulates a special case of a theorem provided in \cite{granstrom_estimation_2012}. Recall that the exponential distribution is a special case of a gamma distribution with shape parameter \mathintext{\alpha=1}.

\noindent\rule{\linewidth}{0.2pt}
\subsubsection*{Lemma 1}
Let \mathintext{p(x)} be a weighted sum of an exponential distribution \mathintext{\mathcal{E}(x;\gamma)} and a gamma distribution \mathintext{\mathcal{G}(x;\alpha,\beta)}:
\begin{equation}
    p(x) = w_1\mathcal{E}(x;\gamma) + w_2\mathcal{G}(x;\alpha,\beta) \;,
\end{equation}
with \mathintext{w_1} and \mathintext{w_2} being the corresponding weights of the mixture and \mathintext{w_1+w_2=1}. Let
\begin{equation}
    q(x) = \mathcal{G}(x;a,b)
\end{equation} be the gamma distribution that minimises the KLD \cite{bishop_pattern_2006} between \mathintext{p(x)} and \mathintext{q(x)} such that:
\begin{equation}
    q^{*}(x) = \text{argmin} \ \text{KL}\left(p(x)||q(x)\right) \;.
\end{equation}
Then the parameters \mathintext{a}  and \mathintext{b} of the gamma distribution \mathintext{q^{*}(x)} are given by the solution to:
\begin{dmath}
\label{eq:gamma_merge_a}
    0 = \log a - \psi_0(a) - \log\left( w_1\frac{1}{\gamma} + w_2\frac{\alpha}{\beta} \right) + w_1(\psi_0(1)-\log \gamma) + w_2(\psi_0(\alpha)-\log \beta) \;,
\end{dmath}
where \mathintext{\psi_0} is the first derivative of the natural logarithm of the gamma function, namely the polygamma function of order 0, and by:
\begin{equation}
\label{eq:gamma_merge_b}
    b = \frac{a}{w_1\frac{1}{\gamma} + w_2\frac{\alpha}{\beta}} \;.
\end{equation}
A proof of this lemma is provided in \cite{granstrom_estimation_2012}.
\noindent\rule{\linewidth}{0.2pt}
\\

After merging the two conditional densities, weighted by their corresponding probability of existence, we obtain a single gamma distribution which is the closest to the mixture in terms of the KLD and we can therefore retain the closed-form solution as the gamma distribution is conjugate to the Poisson distribution.

\begin{figure}[tbh]
    \centering
    \includegraphics[width=\linewidth]{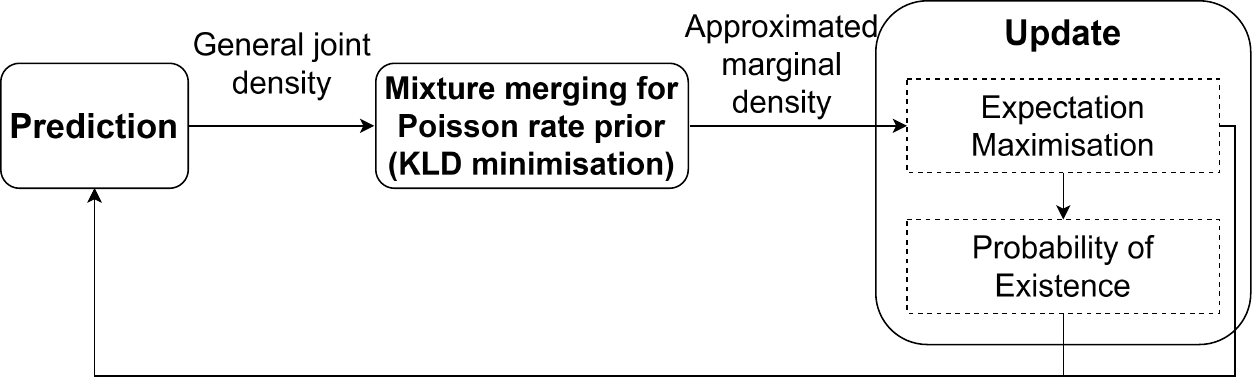}
    \caption{Block diagram of the IE-PHPMHT filter. The general joint density (\ref{eq:p_comp_existence_1}) after the prediction is approximated via a KLD minimisation based mixture merging of the Poisson rates (\ref{eq:prior_lambda_ex}). In the update step, the filter obtains MAP estimates of the target states and Poisson rates via EM and recovers the probability of existence for each potential target for the next multi-Bernoulli prediction.}
    \label{fig:block_diagram}
\end{figure}

The mixture merging is not part of the EM algorithm. It is only performed once every time step to approximate the marginal complete likelihood before applying EM to estimate the states and the Poisson rates. This is illustrated in Fig. \ref{fig:block_diagram}.

\section{PHPMHT with Integrated Probability of Target Existence}
\label{sec:IE-HPMHT}
This section highlights the extension made to impose the targets' existence into the PHPMHT framework and presents its derivation via the expectation step in Section \ref{sec:HPMHT_expectation} and the maximisation step in Section \ref{sec:HPMHT_maximisation} for the IE-PHPMHT. In addition, Section \ref{sec:HPMHT_probEx} shows how the probability of existence can be recovered from the estimated Poisson measurement rates. 
\subsection{Expectation}
\label{sec:HPMHT_expectation}
The expectation step is the first step in order to perform EM and consists essentially of the derivation of the EM auxiliary function; the expectation of the logarithm of the complete data likelihood over the conditional missing data. The expression for the complete data likelihood \mathintext{p_{\text{comp}}} is the joint probability of all variables in the model and is given in (\ref{eq:p_comp_existence_2}). 

Returning to the auxiliary function in (\ref{eq:aux}), by substituting  (\ref{eq:lh_measurements}), (\ref{eq:lh_state}), (\ref{eq:lh_Prate}), and (\ref{eq:lh_assignments_complete}) into (\ref{eq:p_comp_existence_2_expand}), the complete data likelihood is given as:
\begin{align}
\label{eq:p_comp_complete}
    &p_{\text{comp}}\left(N_\timeind, \mathbb{Y}_\timeind, X_\timeind, \Lambda_\timeind \right) \nonumber \\ 
    &= \makesubandsuper{\prod}{\targetind=0}{\Targetind}  p(\makesubandsuper{\vect{x}}{\timeind}{\targetind}) p(\makesubandsuper{\lambda}{\timeind}{\targetind}) 
    \makesubandsuper{\prod}{\pixelindx=1}{\Pixelindx} \text{e}^{-\blue{\makesubandsuper{\Bar{\nu}}{\timeind}{\pixelind}}} \makesubandsuper{\prod}{r=1}{\makesubandsuper{n}{\timeind}{\pixelind}} \makesubandsuper{\lambda}{\timeind}{\makesubandsuper{m}{\timeind}{\pixelind,r}} p^{i,\makesubandsuper{m}{\timeind}{\pixelind,r}}\left( \makesubandsuper{\vect{y}}{\timeind}{\pixelind,r}|\makesubandsuper{\vect{x}}{\timeind}{\makesubandsuper{m}{\timeind}{\pixelind,r}}\right) .
\end{align}

Combining (\ref{eq:p_comp_complete}) and (\ref{eq:lh_measurements}) and inserting it back into (\ref{eq:aux}) results in the final full expression of the auxiliary function:
\small
\begin{align}
\label{eq:aux_comp_unsolved}
    \makesubandsuper{\aux}{\timeind}{\itind} & \left(X_\timeind,\Lambda_\timeind|\hat{X}_\timeind^{(\itind-1)},\hat{\Lambda}_\timeind^{(\itind-1)}\right) \nonumber\\
    =& \makesubandsuper{\sum}{\targetind=0}{\Targetind} \Big[ \log \left\{ p(\makesubandsuper{\vect{x}}{\timeind}{\targetind}) \right\}
    - \makesubandsuper{\lambda}{\timeind}{\targetind} + \log \left\{ p(\makesubandsuper{\lambda}{\timeind}{\targetind}\blue{)}\right\} \Big] \nonumber \\
    &+ \int \makesubandsuper{\sum}{\pixelindx=1}{\Pixelindx} \makesubandsuper{\sum}{r=1}{\makesubandsuper{n}{\timeind}{\pixelind}} \biggl( \log \left\{ \makesubandsuper{\lambda}{\timeind}{\makesubandsuper{m}{\timeind}{\pixelind,r}} \right\}
    + \log \left\{ p^{i,\makesubandsuper{m}{\timeind}{\pixelind,r}}\left( \makesubandsuper{\vect{y}}{\timeind}{\pixelind,r}|\makesubandsuper{\vect{x}}{\timeind}{\makesubandsuper{m}{\timeind}{\pixelind,r}}\right) \right\} \biggr) \nonumber \\
    &\times \makesubandsuper{\prod}{\pixelindx=1}{\Pixelindx} n_\timeind^{\pixelind}! \prod_{r=1}^{n_\timeind^{\pixelind}} \frac{ \, \makesubandsuper{\hat{\lambda}}{\timeind}{\makesubandsuper{m}{\timeind}{\pixelind,r},{(\itind-1)}} \makesubandsuper{p}{}{i,\makesubandsuper{m}{\timeind}{\pixelind,r}}\left( \makesubandsuper{\vect{y}}{\timeind}{\pixelind,r} | \makesubandsuper{\hat{\vect{x}}}{\timeind}{\makesubandsuper{m}{\timeind}{\pixelind,r},{(\itind-1)}} \right)}{\blue{\makesubandsuper{\hat{\Bar{\nu}}}{\timeind}{\pixelind,{(\itind-1)}}}}
    \ \delta \mathbb{Y}_\timeind \;
\end{align}
\normalsize
\blue{where \mathintext{\blue{\hat{\Bar{\nu}}_\timeind^{\pixelind,(\itind-1)}}} is the estimated total number of measurements, which implicitly depends on \mathintext{\blue{\hat{X}_\timeind^{(\itind-1)}}} and \mathintext{\blue{\hat{\Lambda}_\timeind^{(\itind-1)}}}, in cell \mathintext{\blue{i}} from the previous iteration \mathintext{\blue{\itind-1)}} and is given by:} 
\everymath{\color{black}}
\begin{align}
\label{eq:nu_bar_hat}
    \hat{\Bar{\nu}}_\timeind^{\pixelind,(\itind-1)} = \makesubandsuper{\sum}{s=0}{\Targetind} \makesubandsuper{\hat{\lambda}}{\timeind}{s,(\itind-1)} \int_{A^\pixelind} p^{i,s}\left(y|\makesubandsuper{\hat{x}}{\timeind}{s,(\itind-1)}\right)\, dy \;.
\end{align}
Note that the terms in the first line in \eqref{eq:aux_comp_unsolved} have been taken out of the conditional expectation as the states, and Poisson rates do not depend on either the assignments or measurement locations.

\everymath{\color{black}}
\blue{By isolating the components of the auxiliary function in \eqref{eq:aux_comp_unsolved} that depend exclusively on the Poisson rates and target states, and subsequently solving the sequence of set integrals, we obtain two distinct auxiliary functions. This result is formalised in Lemma 2.}

\everymath{\color{black}}
\needspace{4\baselineskip}
\noindent\rule{\linewidth}{0.2pt}
\subsubsection*{\blue{Lemma 2}}
\blue{For the considered models (\ref{eq:lh_measurements}), (\ref{eq:lh_state}), (\ref{eq:lh_Prate}), and (\ref{eq:lh_assignments_complete}), the EM auxiliary function in \eqref{eq:aux}, also given by \eqref{eq:aux_comp_unsolved}, can be decomposed as:}
\begin{align}
\label{eq:aux_div_general}
    \makesubandsuper{\aux}{\timeind}{\itind} = \makesubandsuper{\aux}{\timeind, \Lambda}{\itind} + \makesubandsuper{\aux}{\timeind, X}{\itind} \,,
\end{align}
\blue{where the auxiliary function for the Poisson rates is}
\begin{align}
\label{eq:aux_Lambda}
    \makesubandsuper{\aux}{\timeind,\Lambda}{\itind} & \left(X_\timeind,\Lambda_\timeind|\hat{X}_\timeind^{(\itind-1)},\hat{\Lambda}_\timeind^{(\itind-1)}\right) \nonumber\\
    =& \makesubandsuper{\sum}{\targetind=0}{\Targetind} \Biggl[ \log \left\{ p(\makesubandsuper{\lambda}{\timeind}{\targetind}) \right\} -\makesubandsuper{\lambda}{\timeind}{\targetind} + \log \left\{ \makesubandsuper{\lambda}{\timeind}{\targetind} \right\} \nonumber \\
    &\times \makesubandsuper{\sum}{\pixelind=1}{\Pixelindx} \makesubandsuper{n}{\timeind}{\pixelind}
    \frac{\makesubandsuper{\hat{\lambda}}{\timeind}{\targetind,(\itind-1)}  \int_{A^\pixelind} p^{i,\targetind}\left( y|\makesubandsuper{\hat{\vect{x}}}{\timeind}{\targetind,(\itind-1)} \right) dy}{\makesubandsuper{\hat{\bar{\nu}}}{\timeind}{\pixelind,(\itind-1)}} \Biggr] \\
    =& \makesubandsuper{\sum}{\targetind=0}{\Targetind} \left[ \log \left\{ p(\makesubandsuper{\lambda}{\timeind}{\targetind}) \right\} 
    + \log \left\{ \text{e}^{-\makesubandsuper{\lambda}{\timeind}{\targetind}} {\left(\makesubandsuper{\lambda}{\timeind}{\targetind}\right)}^{\makesubandsuper{\Bar{n}}{\timeind}{\targetind}} \right\} \right] \;,
\end{align}
\blue{with}
\begin{align}
    \makesubandsuper{\Bar{n}}{\timeind}{\targetind} = \makesubandsuper{\sum}{\pixelind=1}{\Pixelind} \makesubandsuper{n}{\timeind}{\pixelind}
    \frac{\makesubandsuper{\hat{\lambda}}{\timeind}{\targetind,(\itind-1)}  \int_{A^\pixelind} p^{i,\targetind}\left( y|\makesubandsuper{\hat{\vect{x}}}{\timeind}{\targetind,(\itind-1)} \right) dy}{\makesubandsuper{\sum}{s=0}{\Targetind} \makesubandsuper{\hat{\lambda}}{\timeind}{s,(\itind-1)} \int_{A^\pixelind} p^{i,s}\left( y|\makesubandsuper{\hat{\vect{x}}}{\timeind}{s,(\itind-1)} \right) dy} \;,
\end{align}
\blue{and the auxiliary function for the target states is:}
\begin{align}
\label{eq:aux_states}
    \makesubandsuper{\aux}{\timeind,X}{\itind} & \left(X_\timeind,\Lambda_\timeind|\hat{X}_\timeind^{(\itind-1)},\hat{\Lambda}_\timeind^{(\itind-1)}\right) \nonumber\\
    =& \makesubandsuper{\sum}{\targetind=0}{\Targetind} \Biggl[ \log \left\{ p(\makesubandsuper{\vect{x}}{\timeind}{\targetind}) \right\} + \makesubandsuper{\hat{\lambda}}{\timeind}{\targetind,(\itind-1)}
    \makesubandsuper{\sum}{\pixelind=1}{\Pixelindx} \frac{\makesubandsuper{n}{\timeind}{\pixelind}}{\makesubandsuper{\hat{\Bar{\nu}}}{\timeind}{\pixelind,(\itind-1)}} \nonumber \\
    & \times \int_{A^\pixelind} \log \left\{ p^{i,\targetind} \left( y|\makesubandsuper{\vect{x}}{\timeind}{\targetind} \right) \right\} p^{i,\targetind} \left( y|\makesubandsuper{\hat{\vect{x}}}{\timeind}{\targetind,(\itind-1)} \right) d y \Biggr] \;.
\end{align}
\blue{The proof of this lemma is provided in the Appendix.}
\noindent\rule{\linewidth}{0.2pt}
\everymath{\color{black}}

In the last line of \eqref{eq:aux_Lambda}, the entire sum over all resolution cells which gives a mean number of measurements from a target has been abbreviated with \mathintext{\makesubandsuper{\Bar{n}}{\timeind}{\targetind}}. 
\subsection{Maximisation}
\label{sec:HPMHT_maximisation}
To complete EM, the EM auxiliary function (\ref{eq:aux_div_general}) has to be optimised as the second step. Since the auxiliary function for the states has not changed in our new formulation, the function of this structure can be maximised using an appropriate point measurement MAP estimator as pointed out in \cite{streit_tracking_2000}. In the case of a linear Gaussian density \mathintext{p^m(y|\makesubandsuper{\vect{x}}{\timeind}{\targetind})} as well as linear Gaussian target dynamics, the maximisation can be done by means of a \textit{Kalman filter} (KF). For other applications, non-linear state estimation techniques such as extended KF or particle filter can be employed as shown in \cite{davey_histogram_2011}.

Since the EM algorithm typically only provides point estimates rather than the full posterior distributions, we restrict our implementation in this paper to applications using a KF. In KF implementations, both the mean and covariance are used to propagate the state density over time. However, it is important to note that in this case the covariance does not necessarily reflect the true mean square error and it has been shown that in the PMHT framework, KFs can tend to underestimate the actual error \cite{willett_pmht_2002}. In more general cases, for the full posterior estimation, techniques such as those used in \cite{bernardo_variational_2003}, may be more appropriate.

In order to estimate the Poisson measurement rates, the \mathintext{\makesubandsuper{\aux}{\timeind, \Lambda}{\itind}} part of the auxiliary function needs to be maximised. Rather than maximising this function directly, we can also find the Poisson rate that maximises that function by finding the mode of the posterior. As it can be seen from (\ref{eq:aux_Lambda}), the right-hand side part is essentially in the form of a Poisson distribution except for the missing normalisation constant in the denominator. However, as this is merely a scaling factor it does not affect the maximisation problem. Since the Poisson rate prior is a gamma distribution, due to conjugacy we get a closed-form expression for the posterior and hence for the EM maximisation in the update. 
The posterior of the Poisson measurement rate is given by:
\begin{align}
\label{eq:lambda_posterior}
    p(\makesubandsuper{\lambda}{\timeind}{\targetind}|\makesubandsuper{\Bar{n}}{\timeind}{\targetind}) \propto p(\makesubandsuper{\lambda}{\timeind}{\targetind}) p(\makesubandsuper{\Bar{n}}{\timeind}{\targetind}|\makesubandsuper{\lambda}{\timeind}{\targetind}) \;,
\end{align}
and it can be shown that due to conjugacy this will again be a gamma distribution. \blue{That is,}
\everymath{\color{black}}
\begin{align}
    p(\makesubandsuper{\lambda}{\timeind}{\targetind}) p(\makesubandsuper{\Bar{n}}{\timeind}{\targetind}|\makesubandsuper{\lambda}{\timeind}{\targetind})
    &= \mathcal{G}(\makesubandsuper{\lambda}{\timeind}{\targetind};\makesubandsuper{\alpha}{\timeind}{\targetind},\makesubandsuper{\beta}{\timeind}{\targetind}) \mathcal{P}(\makesubandsuper{\Bar{n}}{\timeind}{\targetind}|\makesubandsuper{\lambda}{\timeind}{\targetind}) \nonumber \\
    &= \mathcal{G}(\makesubandsuper{\lambda}{\timeind}{\targetind};\makesubandsuper{\alpha}{\timeind}{\targetind} + \makesubandsuper{\Bar{n}}{\timeind}{\targetind},\makesubandsuper{\beta}{\timeind}{\targetind}+1) \nonumber \\
    &= \mathcal{G}(\makesubandsuper{\lambda}{\timeind}{\targetind};\makesubandsuper{a}{\timeind}{\targetind},\makesubandsuper{b}{\timeind}{\targetind}) \;,
\end{align}
\blue{where \mathintext{\mathcal{G(\cdot)}} denotes a gamma distribution and \mathintext{\mathcal{P}(\cdot)} denotes a Poisson distribution.Therefore, the expression in \eqref{eq:aux_Lambda} is maximised, for each \mathintext{\blue{\targetind}}, at the mode of the posterior gamma density:}
\begin{align}
    \makesubandsuper{\hat{\lambda}}{\timeind}{\targetind} &= \underset{\makesubandsuper{\lambda}{\timeind}{\targetind}}{\text{argmax}}\ p(\makesubandsuper{\lambda}{\timeind}{\targetind}) p(\makesubandsuper{\Bar{n}}{\timeind}{\targetind}|\makesubandsuper{\lambda}{\timeind}{\targetind}) \nonumber \\
    &= \underset{\makesubandsuper{\lambda}{\timeind}{\targetind}}{\text{argmax}}\ p(\makesubandsuper{\lambda}{\timeind}{\targetind}|\makesubandsuper{\Bar{n}}{\timeind}{\targetind})
    = \frac{\makesubandsuper{a}{\timeind}{\targetind}-1}{\makesubandsuper{b}{\timeind}{\targetind}} \;,
\end{align}
\blue{where \mathintext{\makesubandsuper{a}{\timeind}{\targetind}} and \mathintext{\makesubandsuper{b}{\timeind}{\targetind}} are the parameters of the posterior gamma distribution in \eqref{eq:lambda_posterior} and the EM iteration index has been omitted for brevity.}

\everymath{\color{black}}
Note that the prior of the Poisson rate is in the form of a gamma distribution even after the marginalisation of the existence due to the proposed KLD minimisation-based mixture merging and hence preserves a closed-form solution.
\subsection{Calculating the Probability of Existence}
\label{sec:HPMHT_probEx}
One of the key points addressed in this paper is to handle an unknown and time-varying number of targets. Now having incorporated the existence paradigm into the PHPMHT framework we have gained a measure that can perform integrated track management. 

The probability of existence \mathintext{\makesubandsuper{r}{\timeind}{\targetind} = p(\makesubandsuper{e}{\timeind}{\targetind}=1|\makesubandsuper{\lambda}{\timeind}{\targetind})} for each potential target is calculated by applying Bayes' rule to (\ref{eq:prior_lambda}):
\begin{align}
\label{eq:calc_prob_ex}
    \makesubandsuper{r}{\timeind}{\targetind} &= \frac{p(\makesubandsuper{\lambda}{\timeind}{\targetind}|\makesubandsuper{e}{\timeind}{\targetind}=1)p(\makesubandsuper{e}{\timeind}{\targetind}=1)}{p(\makesubandsuper{\lambda}{\timeind}{\targetind}|\makesubandsuper{e}{\timeind}{\targetind}=1)p(\makesubandsuper{e}{\timeind}{\targetind}=1) + p(\makesubandsuper{\lambda}{\timeind}{\targetind}|\makesubandsuper{e}{\timeind}{\targetind}=0)p(\makesubandsuper{e}{\timeind}{\targetind}=0)} \nonumber\\
    &= \frac{\mathcal{G}(\makesubandsuper{\lambda}{\timeind}{\targetind};\makesubandsuper{\alpha}{\timeind}{\targetind},\makesubandsuper{\beta}{\timeind}{\targetind})\makesubandsuper{r}{\timeind|\timeind-1}{\targetind}}{\mathcal{G}(\makesubandsuper{\lambda}{\timeind}{\targetind};\makesubandsuper{\alpha}{\timeind}{\targetind},\makesubandsuper{\beta}{\timeind}{\targetind})\makesubandsuper{r}{\timeind|\timeind-1}{\targetind} + \mathcal{E}(\makesubandsuper{\lambda}{\timeind}{\targetind};\makesubandsuper{\gamma}{\timeind}{\targetind})(1-\makesubandsuper{r}{\timeind|\timeind-1}{\targetind})} \;.
\end{align}
To obtain \mathintext{\makesubandsuper{r}{\timeind}{\targetind}} we evaluate the above expression at \mathintext{\makesubandsuper{\hat{\lambda}}{\timeind}{\targetind}} that results from the preceding EM together with the corresponding shape and rate parameters \mathintext{\makesubandsuper{\alpha}{\timeind}{\targetind}}, \mathintext{\makesubandsuper{\beta}{\timeind}{\targetind}}, \mathintext{\makesubandsuper{\gamma}{\timeind}{\targetind}} of the gamma and exponential distribution, and \mathintext{r_{\timeind|\timeind-1}^\targetind} is the predicted probability of existence.

\begin{algorithm}
\caption{IE-PHPMHT}
\begin{algorithmic}[1]
\label{alg:HPMHT_existence_pseudo}
\STATE \textbf{Input:} Measurement $Z_\timeind$.
\STATE \textbf{Output:} Estimated target state $\hat{\vect{x}}_\timeind^\targetind$ and probability of existence $r_k^m$ for \mathintext{m = \{1,...,M_k\}}.
\vspace{.2cm}
\STATE \textbf{Prediction:}
\STATE Predict the state and the existence for surviving targets using (\ref{eq:state_pred}) and (\ref{eq:exist_pred}).
\STATE Approximate (\ref{eq:prior_lambda_ex}) by merging the conditional probabilities (\ref{eq:p_cond_l_e1}) and (\ref{eq:p_cond_l_e0}) into a single gamma distribution using (\ref{eq:gamma_merge_a}) and (\ref{eq:gamma_merge_b}).
\STATE Add \mathintext{M_k^b} new targets, see Sec. \ref{sec:Problem_dynamics}.
\STATE \textbf{Expectation Maximisation:}
\WHILE{not converged}
    \STATE \textbf{Update:}
    \STATE Calculate the auxiliary functions (\ref{eq:aux_Lambda}) and (\ref{eq:aux_states}) using (\ref{eq:quant_aux_lim}) and the previous estimates.
    \STATE Estimate the Poisson rates $\hat{\lambda}_k^m$ and the states $\hat{\vect{x}}^m_{k}$ maximising (\ref{eq:aux_Lambda}) and (\ref{eq:aux_states}) using (\ref{eq:quant_aux_lim}).
\ENDWHILE
\STATE Calculate the probability of existence $r_k^m$ using (\ref{eq:calc_prob_ex}).
\STATE Remove targets with existence probabilities below the termination threshold.
\end{algorithmic}
\end{algorithm}

\subsection{H-PMHT Quantisation for Track-Before-Detect}
The preceding derivations have assumed that we are given count data in the form of \mathintext{N_\timeind~=~\left[ \makesubandsuper{n}{\timeind}{1},..., \makesubandsuper{n}{\timeind}{\pixelind},...,\makesubandsuper{n}{\timeind}{\Pixelind} \right]^\transp} where every element contains an integer that is the number of measurements \mathintext{\makesubandsuper{n}{\timeind}{\pixelind}} that are located inside a resolution cell. However, for the application to TBD multi-target tracking, we apply, the within the H-PMHT framework commonly used quantisation step \cite{streit_tracking_2000}, which allows for the use of intensity data as input measurements. Hence, our measurements are \mathintext{Z_{\timeind}~=~\left[ \makesubandsuper{z}{\timeind}{1}, ...,\makesubandsuper{z}{\timeind}{\pixelind},..., \makesubandsuper{z}{\timeind}{\Pixelind} \right]^\transp} where each element represents an intensity value in a certain resolution cell. 

For the quantisation, it is assumed that the intensity values in every resolution cell are arbitrarily quantised \mathintext{\makesubandsuper{n}{\timeind}{\pixelind} = \lfloor \frac{\makesubandsuper{z}{\timeind}{\pixelind}}{\hbar^2} \rfloor}. To remove the quantisation from the actual implementation and apply the algorithm directly to intensity data, the limiting case \mathintext{\hbar^2 \rightarrow 0} of the arbitrary quanta is considered. The number of quantised measurements can also be expressed as \mathintext{\makesubandsuper{n}{\timeind}{\pixelind} = \lfloor \frac{\makesubandsuper{z}{\timeind}{\pixelind}}{\hbar^2} \rfloor = \frac{\makesubandsuper{z}{\timeind}{\pixelind}}{\hbar^2} - \epsilon}, where \mathintext{0\leq\epsilon<\hbar^2}. With this we get
\begin{align}
    \lim_{\hbar^2 \to 0} \hbar^2\makesubandsuper{n}{\timeind}{\pixelind} &= \lim_{\hbar^2 \to 0} (\makesubandsuper{z}{\timeind}{\pixelind}-\hbar^2\epsilon) = \makesubandsuper{z}{\timeind}{\pixelind} \;,
\end{align}
and
\begin{align}
\lim_{\hbar^2 \to 0} \hbar^2||\makesubandsuper{N}{\timeind}{}|| &= \makesubandsuper{\sum}{\pixelindx=1}{\Pixelindx} \lim_{\hbar^2 \to 0} \hbar^2\makesubandsuper{n}{\timeind}{\pixelind}\\ \nonumber
    &= \makesubandsuper{\sum}{\pixelind=1}{\Pixelind} \makesubandsuper{z}{\timeind}{\pixelind} = ||\makesubandsuper{Z}{\timeind}{}|| \;.
\end{align}
This way of recovering the intensity data can be further applied to the EM auxiliary function
\begin{equation}
\label{eq:quant_aux_lim}
    \lim_{\hbar^2 \to 0} \hbar^2 \makesubandsuper{\aux}{\timeind}{\itind} = \lim_{\hbar^2 \to 0} \hbar^2 \makesubandsuper{\aux}{\timeind, \Lambda}{\itind} + \lim_{\hbar^2 \to 0} \hbar^2 \makesubandsuper{\aux}{\timeind, X}{\itind} \,.
\end{equation}

As a result, the filter can be applied directly to intensity data without any quantisation in the actual implementation. For a detailed discussion on the quantisation the reader is referred to \cite{gaetjens_tale_2021}. A summary of the proposed IE-PHPMHT is provided in Algorithm \ref{alg:HPMHT_existence_pseudo}.

\section{Simulation Results}
\label{sec:SimulationResults}
In this section, we evaluate the performance of the proposed IE-PHPMHT formulation, which incorporates existence probabilities, by comparing it with other state-of the-art methods in two distinct multi-target TBD scenarios. Section \ref{sec:Sim_scen1_setup} and \ref{sec:Sim_scen2_setup} outline the simulation setup for each scenario including the measurement and dynamic models as well as the algorithms used for comparison. In Section \ref{sec:Sim_scen1_results} and \ref{sec:Sim_scen2_results} we present and discuss the results for each scenario, respectively.
\subsection{Scenario 1 }
\label{sec:SimRes_setup}
\begin{figure}[b]
    \centering
    \includegraphics[width=\linewidth]{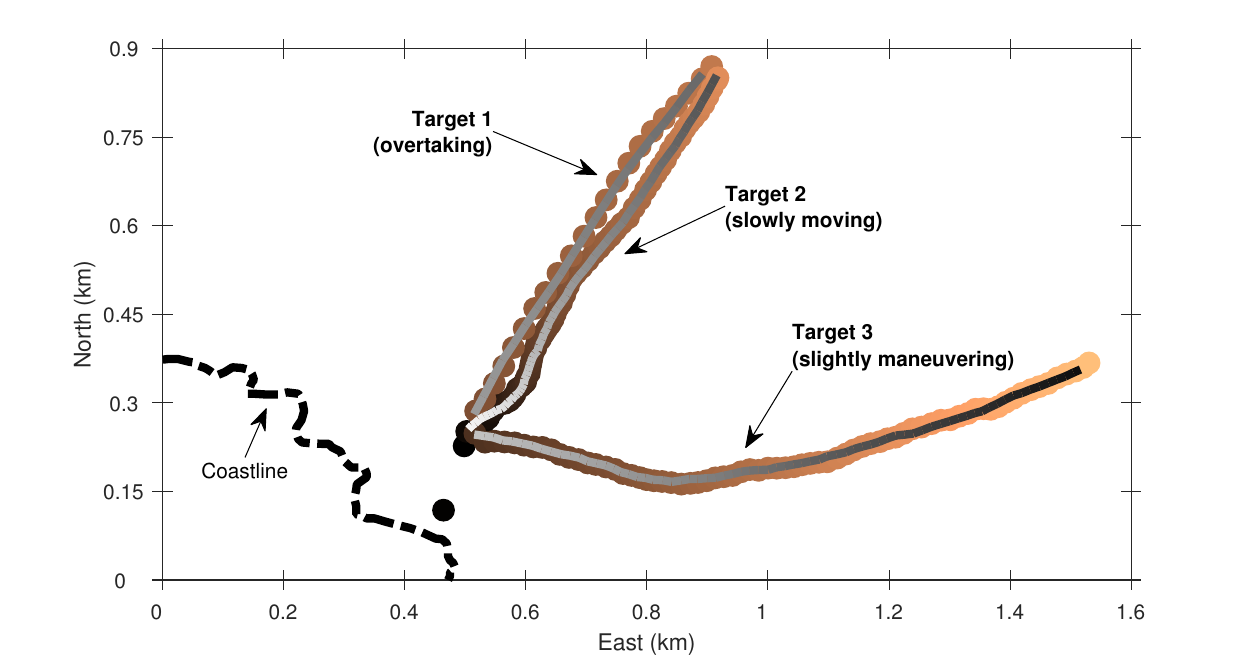}
    \caption{Illustration of the simulation scenario, shown with one realisation of the IE-PHPMHT tracking results. The straight lines indicate the ground truth and the dots are the estimated trajectories of the IE-PHPMHT at each time step where the colour fade illustrates the evolution over time.}
    \label{fig:Scen1_trajectories}
\end{figure}
The first scenario is designed to showcase challenges that are faced in maritime radar tracking scenarios. Suppose a surveillance region that covers an area of 4 kilometres in East and 1.5 kilometres in North direction. The entire image constitutes 40\,000 resolution cells, divided into 400 \mathintext{\times} 100 cells with a resolution of 10 and 15 metre, respectively. The scenario has up to 3 targets that appear and disappear at different times to test the performance of handling a time-varying and unknown number of targets. The trajectories are also designed to test multi-target tracking performance in a way that two targets are in close proximity to each other for most of the time steps.

In a real-world application, this could represent a pier where targets are likely to depart at different times, and one slower vessel is overtaken by a faster vessel. The third vessel performs a light manoeuvre.

The ground truth trajectories together with one example of the IE-PHPMHT estimates are shown in Fig. \ref{fig:Scen1_trajectories}. 

In real-world applications, the target amplitude of radar returns is typically not constant \cite{richards_principles_2010}. Therefore, we examine two different radar target fluctuation models: the Swerling 0 and Swerling I models \cite{swerling_radar_1997}.

\subsubsection{Simulation Setup}
\label{sec:Sim_scen1_setup}
In our numerical studies, every target in the two-dimensional surveillance area is described by its state vector \mathintext{\makesubandsuper{\vect{x}}{\timeind}{\targetind}~=~\left[ \makesubandsuper{p}{x,\timeind}{m} \ \makesubandsuper{v}{x,\timeind}{m} \ \makesubandsuper{p}{y,\timeind}{m} \ \makesubandsuper{v}{y,\timeind}{m} \right]^{\transp}} where \mathintext{\makesubandsuper{p}{x,\timeind}{m} ,\ \makesubandsuper{p}{y,\timeind}{m}} are the Cartesian coordinates of the target's position, and \mathintext{\makesubandsuper{v}{x,\timeind}{m} ,\ \makesubandsuper{v}{y,\timeind}{m}} are the velocities. Given a target survives from the previous time step to the next, the individual state transition density, which is independent of the other targets, is given by:
\begin{equation}
\label{eq:target_dynamics}
    p(\makesubandsuper{\vect{x}}{\timeind}{\targetind}|\makesubandsuper{\vect{x}}{\timeind-1}{\targetind}) = \Gaussian(\makesubandsuper{\vect{x}}{\timeind}{\targetind}; \vect{F}\makesubandsuper{\vect{x}}{\timeind}{\targetind},\vect{Q})\;,
\end{equation}
where \mathintext{\Gaussian(\makesubandsuper{\vect{x}}{\timeind}{\targetind}; \vect{F}\makesubandsuper{\vect{x}}{\timeind}{\targetind},\vect{Q})} denotes the linear-Gaussian kinematic model assumption. The transition matrix \mathintext{\vect{F}} and the process noise covariance \mathintext{\vect{Q}} are chosen according to a nearly constant velocity (CV) model \cite{bar-shalom_estimation_2001}:
\begin{align}
    \vect{F} = \begin{bmatrix}
                1 & T & 0 & 0 \\
                0 & 1 & 0 & 0 \\
                0 & 0 & 1 & T \\
                0 & 0 & 0 & 1 
               \end{bmatrix}, 
    \vect{Q} = q \begin{bmatrix}
                \frac{1}{3} T^3 & \frac{1}{2} T^2 & 0 & 0 \\
                \frac{1}{2} T^2 & T & 0 & 0 \\
                0 & 0 & \frac{1}{3} T^3 & \frac{1}{2} T^2 \\
                0 & 0 & \frac{1}{2} T^2 & T
               \end{bmatrix}\;,
\end{align}
with the process noise variance \mathintext{q = 0.01\,\text{m}^2/\text{s}^3} and a sample period of \mathintext{T =1\,\text{s}}.

Measurements were created according to the following model. The intensity in every resolution cell is given by the sum of target contributions and noise contributions. Hence, \mathintext{\makesubandsuper{z}{\timeind}{\pixelind}} can be denoted as:
\begin{equation}
    \makesubandsuper{z}{\timeind}{\pixelind}
    = \makesubandsuper{\sum}{\targetind=1}{M} \makesubandsuper{s}{\timeind}{\pixelind,\targetind} + \makesubandsuper{w}{\timeind}{\pixelind} 
    = \makesubandsuper{\sum}{\targetind=1}{M} \makesubandsuper{A}{\timeind}{\targetind} \makesubandsuper{h}{}{\pixelind}(\makesubandsuper{\vect{x}}{\timeind}{\targetind}) + \makesubandsuper{w}{\timeind}{\pixelind} \;,
\end{equation}
where \mathintext{\makesubandsuper{A}{\timeind}{\targetind}} is the amplitude of the measured target and \mathintext{\makesubandsuper{h}{}{\pixelind}(\makesubandsuper{\vect{x}}{\timeind}{\targetind})} defines the influence of each target to the intensity in every resolution cell, which is commonly referred to as the target's \textit{point spread function} (PSF). If no target is present, the intensity is solely determined by the noise component. We used a Gaussian PSF, without the normalisation such that the maximum value is unity, which is given by:
\begin{equation}
    \makesubandsuper{h}{}{\pixelind}(\makesubandsuper{\vect{x}}{\timeind}{\targetind}) = \exp{ \left( -\frac{(\makesubandsuper{c}{x}{\pixelind}-\makesubandsuper{p}{x,\timeind}{\targetind})^2}{2\makesubandsuper{\sigma}{x}{2}} - \frac{(\makesubandsuper{c}{y}{\pixelind}-\makesubandsuper{p}{y,\timeind}{\targetind})^2}{2\makesubandsuper{\sigma}{y}{2}} \right) }\;,
\end{equation}
where \mathintext{\makesubandsuper{\sigma}{x}{2}} and \mathintext{\makesubandsuper{\sigma}{y}{2}} determine the spread in x- and y-direction, and \mathintext{\makesubandsuper{c}{x}{\pixelind}} and \mathintext{\makesubandsuper{c}{y}{\pixelind}} are the centres of each resolution cell, respectively. The noise was assumed to be Rayleigh distributed, which is a commonly used assumption for radar applications, as it results from the magnitude of real- and imaginary zero-mean Gaussian noise parts with variance \mathintext{\sigma_w^2} \cite{richards_fundamentals_2014}. Hence, the noise distribution is given as:
\begin{equation}
    p(\makesubandsuper{w}{\timeind}{\pixelind};\upsilon) = \mathcal{R}(\makesubandsuper{w}{\timeind}{\pixelind};\upsilon)
    = \frac{2\makesubandsuper{w}{\timeind}{\pixelind}}{\upsilon} \exp{\left(-\frac{\left( \makesubandsuper{w}{\timeind}{\pixelind} \right)^2}{\upsilon}\right)} \;,
\end{equation}
with scale parameter \mathintext{\upsilon = 2\sigma_w^2}, and mean noise power \mathintext{\mathbb{E} \left[ \left( \makesubandsuper{w}{\timeind}{\pixelind} \right)^2 \right] = \upsilon}. 

\blue{For a single target, considering the Swerling 0 target plus noise case, the amplitude distribution in each cell is given by a Rician distribution  \cite{mallick_integrated_2013}:}
\everymath{\color{black}}
\begin{align*}
    p_{\mathrm{SW0}}\left(\makesubandsuper{z}{\timeind}{\pixelind};\makesubandsuper{s}{\timeind}{\pixelind},\upsilon\right)
    &= \mathrm{Rice}\left(\makesubandsuper{z}{\timeind}{\pixelind};\makesubandsuper{s}{\timeind}{\pixelind},\upsilon\right)  \nonumber\\
    &=  \frac{2\makesubandsuper{z}{\timeind}{\pixelind}}{\upsilon} \exp{\left( -\frac{\left({\makesubandsuper{z}{\timeind}{\pixelind}}\right)^2 + \left(\makesubandsuper{s}{\timeind}{\pixelind}\right)^2}{\upsilon} \right)}\nonumber \\ 
    & \times\mathcal{I}_0 \left( \frac{2\makesubandsuper{z}{\timeind}{\pixelind} \makesubandsuper{s}{\timeind}{\pixelind}}{\upsilon} \right)\;,
\end{align*}
\blue{where \mathintext{\mathcal{I}_0} is the modified Bessel function of order \mathintext{0}.}
\everymath{\color{black}}
\blue{In the Swerling I case, the target amplitude follows a Rayleigh distribution \cite{lepoutre_multitarget_2016}. Since the sum of two Gaussian random variables is Gaussian, the amplitude in each cell is Rayleigh distributed, but with parameter \mathintext{\blue{\upsilon + \makesubandsuper{s}{\timeind}{\pixelind}}} i.e.,:}

\everymath{\color{black}}
\begin{align*}
    p_{\mathrm{SW1}}\left(\makesubandsuper{z}{\timeind}{\pixelind};\makesubandsuper{s}{\timeind}{\pixelind},\upsilon\right)
    &= \mathcal{R}\left(\makesubandsuper{z}{\timeind}{\pixelind};\upsilon + \makesubandsuper{s}{\timeind}{\pixelind}\right)  \nonumber\\
    &= \frac{2\makesubandsuper{z}{\timeind}{\pixelind}}{\upsilon + \makesubandsuper{s}{\timeind}{\pixelind}} \exp{\left(-\frac{\left( \makesubandsuper{z}{\timeind}{\pixelind} \right)^2}{\upsilon + \makesubandsuper{s}{\timeind}{\pixelind}}\right)} \;.
\end{align*}
\everymath{\color{black}}

\begin{figure}[t]
    \centering
    \includegraphics[width=\linewidth]{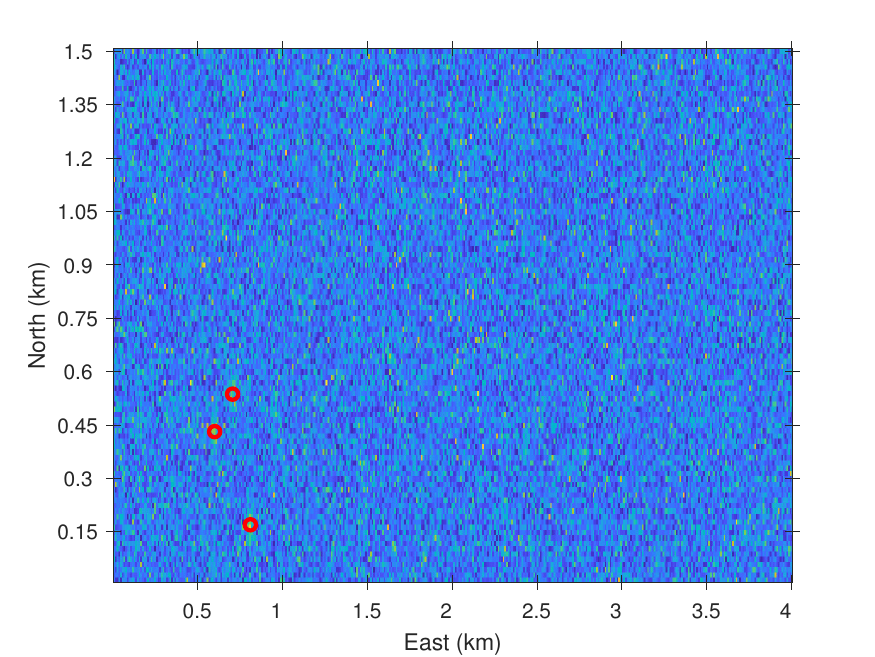}
    \caption{A snapshot of a full radar image at time step k = 45 with 40000 resolution cells covering a surveillance region of 4 km in the east and 1.5 km in the north direction with a target SNR of 5 dB. The red circles show the true target positions.}
    \label{fig:8}
\end{figure}

\begin{table}[htb]
\centering
\caption{Tuning Parameters Overview}
\label{tab:overview_param_scen1}
\renewcommand{\arraystretch}{1.3}
\begin{tabularx}{\columnwidth}{lll}
\hline
\textbf{Quantity} & \textbf{Symbol/[Unit]} & \textbf{Value} \\
\hline
Common Parameters & & \\
\hline
Sample interval & T [s] & 1 \\
Initial position target 1,2,3 & $\vect{p}_0\ \left[ \text{m} \right]$ & $\left[ 500 \ 250 \right]^\transp$ \\
Initial velocity target 1 & $\vect{v}_{0}^1\ [ \text{m}\text{s}^{-1} ]$ & $\left[ 16 \ 30 \right]^\transp$ \\
Initial velocity target 2 & $\vect{v}_{0}^2\ [ \text{m}\text{s}^{-1} ]$ & $\left[ 5 \ 7.5 \right]^\transp$ \\
Initial velocity target 3 & $\vect{v}_{0}^3\ [ \text{m}\text{s}^{-1} ]$ & $\left[ 22.5 \ \text{-}2.5 \right]^\transp$ \\
Process noise intensity & $q\ [ \text{m}^2\text{s}^{-3} ]$ & 0.01 \\
Target spread & $\sigma_x, \ \sigma_y$ [m] & $\sqrt{20}, \ \sqrt{90}$\\
Initial Gamma distr. shape parameter & $\alpha$ [-] & 20 \\
Initial Gamma distr. rate parameter & $\beta$ [-] & 1 \\
\hline
IE-PHPMHT & & \\
\hline
Probability of survival & $p_s$ [-] & 0.98 \\
Probability of birth & $p_b$ [-] & 0.00001 \\
Existence confirmation threshold & $t_c$ [-] & 0.5 \\
Existence termination threshold & $t_d$ [-] & 0.000001 \\
\hline
TH-PHPMHT & & \\
\hline
Exponential forgetting factor Swerling 0 & $\eta_{SW0}$ [-] & 3 \\
Exponential forgetting factor Swerling I & $\eta_{SW1}$ [-] & 10 \\
Track SNR confirmation threshold & $S_c$ [dB] & 0 \\
Track SNR termination threshold & $S_d$ [dB] & -10 \\
\hline
\end{tabularx}
\end{table}

Based on this, for a given target, we define the peak-SNR in dB as:
\begin{equation}
    \text{SNR}
    = 10 \log_{10} \left( \frac{\mathbb{E} \left[ \text{max} \left\{ \left( \makesubandsuper{s}{}{\pixelind} \right)^2 \right\} \right]}
           {\mathbb{E} \left[ \left( \makesubandsuper{w}{}{\pixelind} \right)^2 \right]} \right)
    = 10 \log_{10} \left( \frac{A^2}{\upsilon} \right) \;.
\end{equation}

In the implementation, the densities that determine the influence of each target on each resolution cell in the measurement model, see (\ref{eq:meas_mod_intensity}), were assumed to be Gaussian i.e. \mathintext{p^\targetind(y|\makesubandsuper{x}{\timeind}{\targetind}) = \Gaussian(y; H\makesubandsuper{x}{\timeind}{\targetind}, R)} with \mathintext{H = \left[ \begin{smallmatrix} 1 & 0 & 0 & 0 \\ 0 & 0 & 1 & 0 \end{smallmatrix} \right]} and \mathintext{R = \operatorname{diag}(\left[\makesubandsuper{\sigma}{x}{2}\ \makesubandsuper{\sigma}{y}{2}\right]^\transp)}. The clutter density \mathintext{p^0(y)} was assumed to be a uniform distribution.

Since the basic, core H-PMHT and all available extensions in the literature can not be used in applications with an unknown and time-varying number of targets, the standard approach, e.g. reported in \cite{davey_detecting_2011,davey_comparison_2008,guo_sa-hpmht_2023}, is to use an additional track management part, wrapped around the core tracking filter. These approaches have in common that they base the initialisation, maintenance, and termination of targets on estimated SNRs, which are then tested against predefined thresholds. Therefore, we compared our proposed method with this approach. For the comparison, we based our implementation on \cite{gaetjens_histogram-pmht_2017} for the core tracking part and \cite{mallick_integrated_2013,davey_detecting_2011} for the track management part with the threshold values taken from \cite{davey_comparison_2008}.

In the remainder, our proposed extension of the PHPMHT, which incorporates existence probabilities for integrated track management, is referred to as IE-PHMHT whereas the SNR-based thresholding method is referred to as TH-PHPMHT.

The birth model used permits the birth of at most one target per time step at a known birth location. For the IE-PHPMHT, a target is born with a probability of birth \mathintext{p_b = 0.98} and a Gaussian birth density with the mean \mathintext{\makesubandsuper{\vect{x}}{b}{} = \left[ 500\,\text{m} \ 0 \ 250\,\text{m} \ 0 \right]^\transp} and covariance \mathintext{P_b = \operatorname{diag}(\left[100\,\mathrm{m}^2 \ 10\,\mathrm{m}^2\mathrm{s}^{-2} \ 100\,\mathrm{m}^2 \ 10\,\mathrm{m}^2\mathrm{s}^{-2}\right]^\transp)}. For the TH-PHPMHT, the birth locations were given in the same manner, but, as it does not have any existence probabilities, with just the initial values for the Gamma distribution. The Gamma distributions of both filters were initialised with the same values, see Table \ref{tab:overview_param_scen1}.

Tracks with a probability of existence greater than 0.5 were declared as detected tracks and tracks with a probability of existence less than \mathintext{1 \times 10^{-6}} were deleted. The confirmation and termination threshold for the SNR-based track management part was set to 0 dB and -10 dB, respectively.

The values for the exponential decay factor of the TH-PHPMHT were set to \mathintext{\eta_{SW0} = 3} and \mathintext{\eta_{SW1} = 10} as suggested in \cite{vu_histogram-pmht_2015,gaetjens_histogram-pmht_2017}, which explicitly deals with Swerling 0 and Swerling I target fluctuation models. The simulation ran for 100 time steps, and the targets appeared at time step 
\mathintext{\makesubandsuper{k}{b}{1} = 40}, 
\mathintext{\makesubandsuper{k}{b}{2} = 5}, 
\mathintext{\makesubandsuper{k}{b}{3} = 25},
and disappeared at time step
\mathintext{\makesubandsuper{k}{d}{1} = 60}, 
\mathintext{\makesubandsuper{k}{d}{2} = 70}, 
\mathintext{\makesubandsuper{k}{d}{3} = 95}.

A complete overview of the used parameters is given in Table \ref{tab:overview_param_scen1}.

\subsubsection{Results}
\label{sec:Sim_scen1_results}
\begin{figure}[t]
\centering
\subfloat[RMS GOSPA error]{\includegraphics[width=0.475\columnwidth]{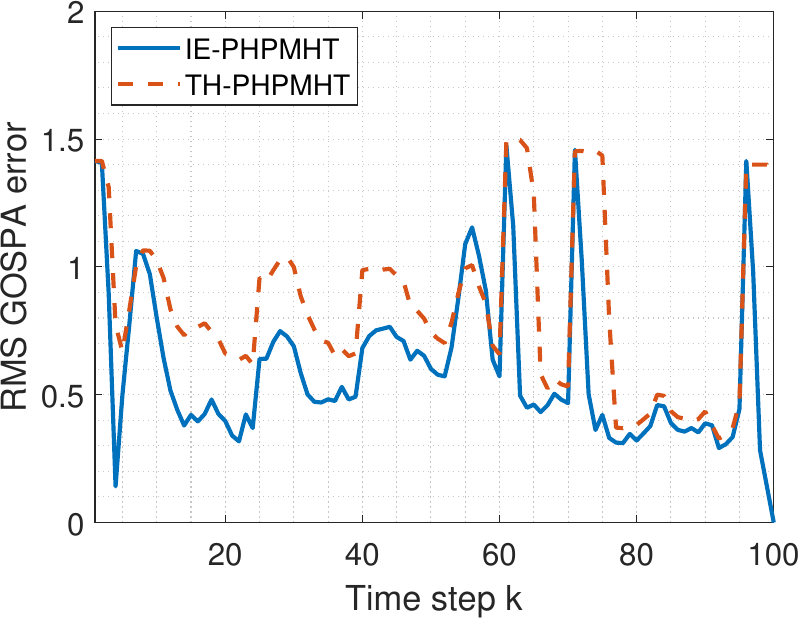}
\label{fig:GOSPA_Scen1_SW0_a}}
\subfloat[RMS GOSPA localisation error]{\includegraphics[width=0.475\columnwidth]{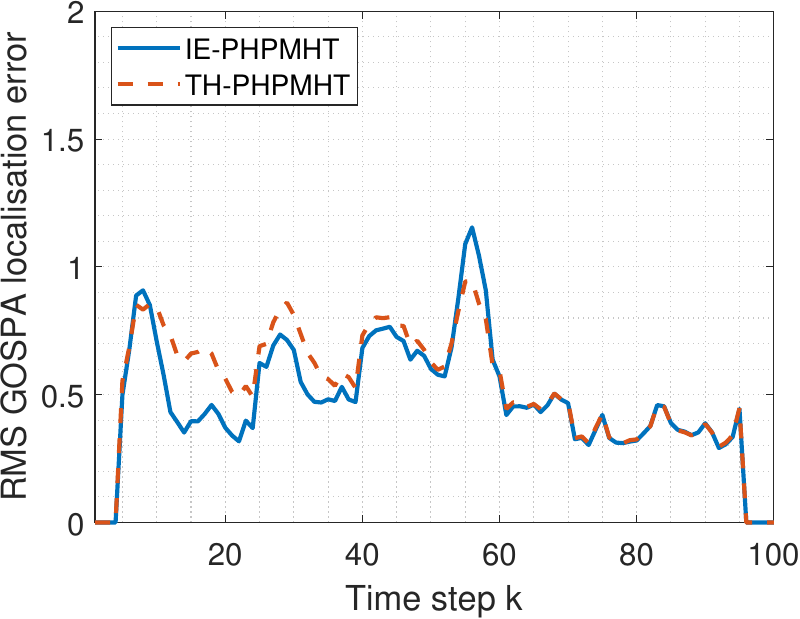}
\label{fig:GOSPA_Scen1_SW0_b}}
\\
\subfloat[RMS GOSPA false target error]{\includegraphics[width=0.475\columnwidth]{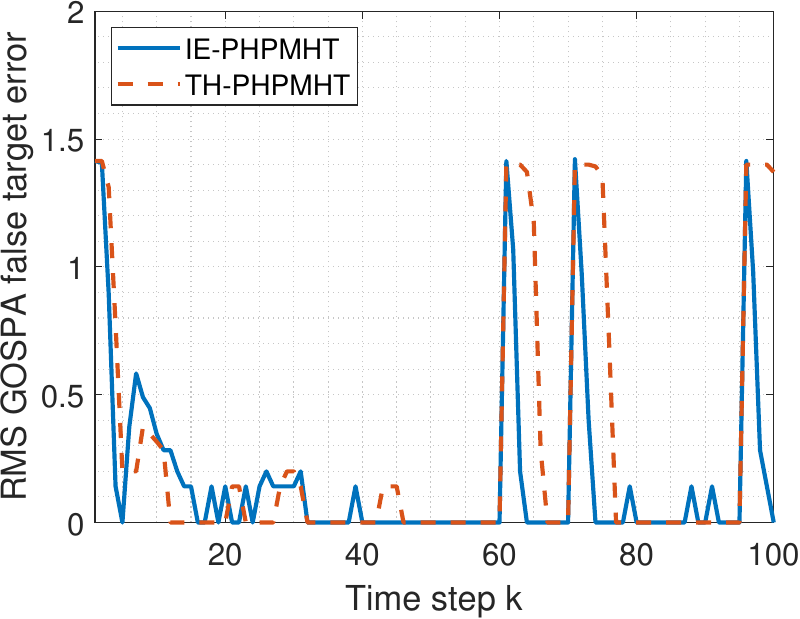}
\label{fig:GOSPA_Scen1_SW0_c}}
\subfloat[RMS GOSPA missed target error]{\includegraphics[width=0.475\columnwidth]{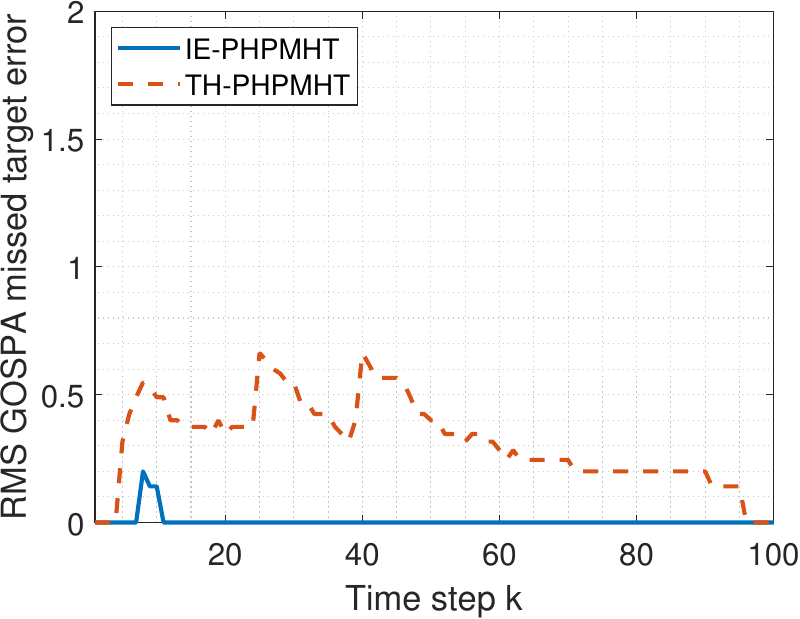}
\label{fig:GOSPA_Scen1_SW0_d}}
\caption{RMS GOSPA errors and their decompositions for an SNR = 5 dB Swerling 0 target averaged over 100 Monte Carlo trials.}
\label{fig:GOSPA_Scen1_SW0}
\end{figure}
The performance assessment and comparison was done by means of the \textit{generalised optimal sub-pattern assignment} (GOSPA) metric \cite{rahmathullah_generalized_2017}. The parameters of the GOSPA metric, the exponent \mathintext{p}, the cardinality factor \mathintext{\alpha}, and the maximum allowable localisation error \mathintext{c} were set to \mathintext{\alpha = 2}, \mathintext{p = 2}, and \mathintext{c = 2L}, where \mathintext{L} represents the dimension of a resolution cell. At every time step, the \textit{root mean square} (RMS) GOSPA for 100 Monte Carlo runs was calculated. Besides the RMS GOSPA, its three individual components, the false target error, the localisation error, and the missed target error were considered. The errors presented are normalised by the dimension of a resolution cell. The RMS GOSPA and its decomposition over time are given in Fig. \ref{fig:GOSPA_Scen1_SW0}.

In Fig. \ref{fig:GOSPA_Scen1_SW0_a} the IE-PHPMHT is shown to have a better performance than the TH-PHPMHT as during the complete simulation run the RMS GOSPA is lower or equal. As expected, the localisation error is essentially the same, which can be seen from Fig \ref{fig:GOSPA_Scen1_SW0_b}, as both methods use the same core algorithm. In both cases, the error is the highest when target 1 overtakes target 2 and they are in close proximity such that their measurements merge (\mathintext{k = 56}). In Fig. \ref{fig:GOSPA_Scen1_SW0_c}, showing the false target error, we see that overall both methods maintain an equally low number of false targets. The only exceptions are when the targets disappear at time steps 60, 70, and 95. The IE-PHPMHT terminates the tracks considerably faster. In Fig. \ref{fig:GOSPA_Scen1_SW0_d} we see that the IE-PHPMHT performs considerably better in terms of track initialisation. The TH-PHPMHT does not only take a longer time when the targets appear it also consistently misses targets over the entire simulation period. The IE-PHPMHT handles the case of the merged measurements well and is still able to separate the targets as there is no noticeable increase in false and missed targets at this time step.

\begin{figure}[t]
\centering
\subfloat[RMS GOSPA error]{\includegraphics[width=0.475\columnwidth]{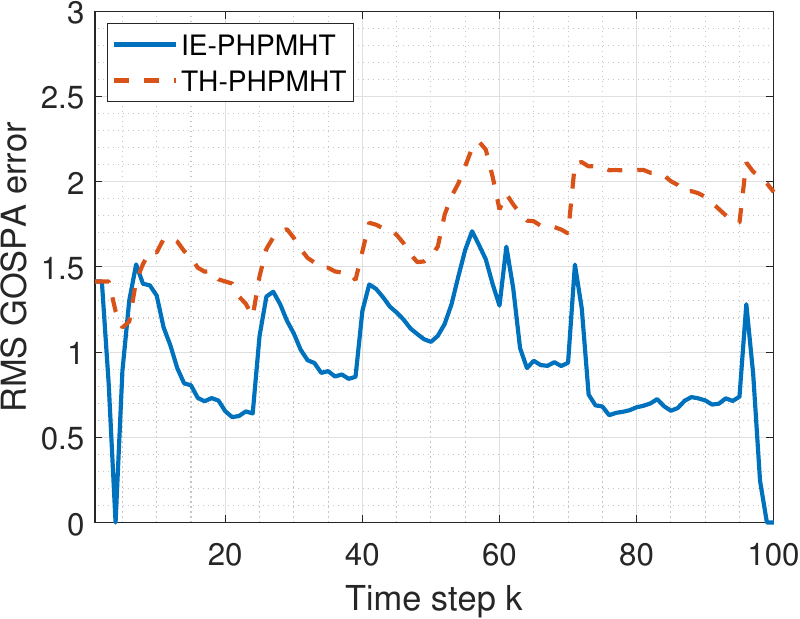}
\label{fig:GOSPA_Scen1_SW1_a}}
\subfloat[RMS GOSPA localisation error]{\includegraphics[width=0.475\columnwidth]{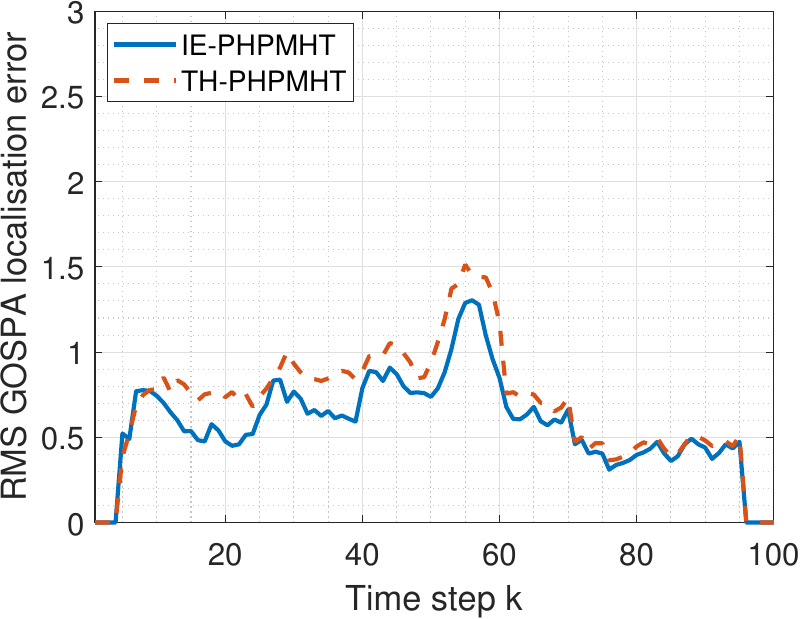}
\label{fig:GOSPA_Scen1_SW1_b}}
\\
\subfloat[RMS GOSPA false target error]{\includegraphics[width=0.475\columnwidth]{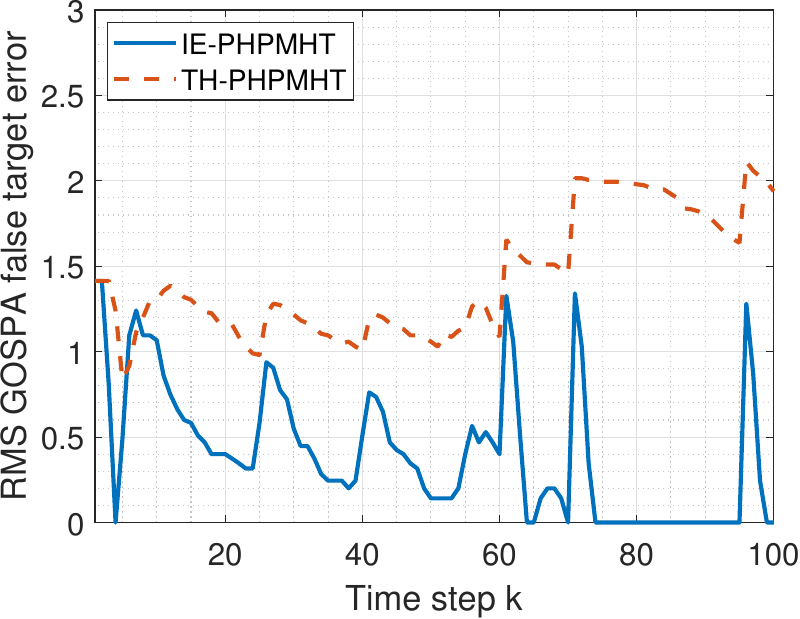}
\label{fig:GOSPA_Scen1_SW1_c}}
\subfloat[RMS GOSPA missed target error]{\includegraphics[width=0.475\columnwidth]{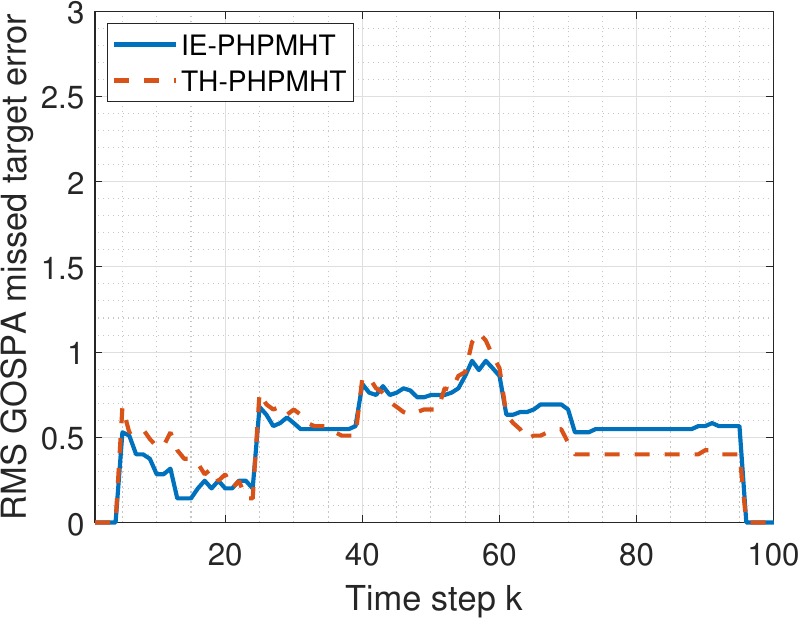}
\label{fig:GOSPA_Scen1_SW1_d}}
\caption{RMS GOSPA errors and their decompositions for an SNR = 12 dB Swerling I target averaged over 100 Monte Carlo trials.}
\label{fig:GOSPA_Scen1_SW1}
\end{figure}

In Fig. \ref{fig:GOSPA_Scen1_SW1} the RMS GOSPA and its decomposition over time is given for the same scenario, but with targets that have amplitude fluctuations following a Swerling I fluctuation model.
In Fig. \ref{fig:GOSPA_Scen1_SW1_a} the total RMS GOSPA error, shown for every time step, indicates that the IE-PHPMHT performs better than the TH-PHPMHT. This is mainly due to the considerably lower number of false targets. Both show peaks during target initialisation, termination, and the overtaking manoeuvre as these are the challenging parts especially \blue{with} a non-constant amplitude. Nevertheless, the IE-PHPMHT adjusts faster towards the real number of targets.
In Fig. \ref{fig:GOSPA_Scen1_SW1_b}, again, both show more or less the same localisation error, with a slightly increased error of the TH-PHPMHT during the initialisation phase.
In Fig. \ref{fig:GOSPA_Scen1_SW1_c} it can be seen that the TH-PHPMHT has a fairly consistent number of false targets. This results in a worse performance when compared with the IE-PHPMHT. 
\blue{The reason is that the existence-based IE-PHPMHT does not include the forgetting factor that propagates the Poisson prior over time to handle target fluctuations, which, in contrast, leads the SNR-based method to retain false targets over several time steps.
}
Fig. \ref{fig:GOSPA_Scen1_SW1_d} shows an increase in missed targets compared to the previous simulation which clearly is caused by the Swerling I fluctuation model as the targets' amplitude can drop significantly for several time steps. In direct comparison, the IE-PHPMHT and the TH-PHPMHT perform almost equally, while the latter has a minimally lower number of false targets after the overtaking manoeuvre. This is likely because the correlation of the Poisson rate prior leads to tracking the targets longer when dealing with amplitude fluctuations.

\begin{table}[t]
\centering
\caption{Scenario 1 Swerling 0 and Swerling I Average RMS GOSPA}
\label{tab:AVG_RMSGOSPA_Scen1}
\renewcommand{\arraystretch}{1.3}
\begin{tabular}{l||c|c|c|c}
\hline
\multicolumn{1}{l||}{\textbf{Method}} & \textbf{Error tot.} & \textbf{Error loc.} & \textbf{Error fal.} & \textbf{Error mis.} \\ \hline
\multicolumn{1}{l||}{\textit{Swerling 0}} & & & & \\ 
IE-PHPMHT & \textbf{0.67} & \underline{0.53} & \underline{0.40} & \underline{0.03} \\ 
TH-PHPMHT & 0.91 & 0.58 & 0.60 & 0.36 \\ \hline
\multicolumn{1}{l||}{\textit{Swerling I}} & & & & \\
IE-PHPMHT & \textbf{1.03} & \underline{0.64} & \underline{0.57} & 0.57 \\ 
TH-PHPMHT & 1.76 & 0.78 & 1.47 & \underline{0.55} \\ \hline
\end{tabular}
\end{table}

As an overall comparison, the total RMS GOPSA as an average across all time steps is shown in Table \ref{tab:AVG_RMSGOSPA_Scen1}. For Swerling 0 targets, the IE-PHPMHT achieves a lower total RMS GOSPA compared to the TH-PHPMHT, primarily due to its lower total localisation error, false target error, and missed target error. Although the localisation errors are similar between the two methods, the TH-PHPMHT misses significantly more targets. Furthermore, the higher false target error suggests that simply adjusting the threshold levels is not sufficient as lowering the threshold to reduce missed targets would result in an increased false target error.

For Swerling I targets, the IE-PHPMHT has a lower total RMS GOSPA error mainly because of a significantly lower total false target error. The total localisation error as well as the total missed target error are comparable. In contrast, the TH-PHPMHT produces a greater total number of false targets. This is likely due to the exponential forgetting factor that leads to a higher correlation between the past estimates on the updated estimates of the Poisson rate. The incentive behind the correlation is to prevent tracks from being terminated when the target amplitude drops for several consecutive time steps \cite{gaetjens_tale_2021} which results in a slightly lower number of the total missed target error. However, this correlation implies also promoting tracks that were initiated by clutter which causes a higher number of false targets.

In summary, this suggests that for both fluctuation models, the IE-PHPMHT model, in which the Poisson rate prior is calculated based on target existence and non-existence, is more balanced in producing false or missed tracks. A key difference between the method in this paper and the Poisson H-PMHT formulation in \cite{gaetjens_histogram-pmht_2017} is the propagation of the Poisson prior over time. From our generative model depicted in Fig. \ref{fig:Bayes_net_model_2} and from (\ref{eq:lh_Prate}) we can see that there is no need for a dynamic model of the Poisson rates as the generation of measurements is only driven by a Bernoulli target that may or may not exist. In addition, this removes the need for the heuristic decay factor in \cite{gaetjens_histogram-pmht_2017} to propagate the Poisson prior over time. Instead, we have introduced an intuitive dynamics that is purely based on the probability of survival of each Bernoulli target and has an interpretable value. Moreover, the probability of survival enables dealing with target fluctuations and leads to better comparability to other tracking algorithms as this is a commonly used way of modelling. Note that the IE-PHPMHT achieved better results for both fluctuation models without changing tuning parameters. In contrast, for the TH-PHPMHT the exponential decay factor for the correlation of the Poisson rates was set to the appropriate fluctuation model. This also suggests that the IE-PHPMHT provides more resilience to target fluctuations even though it does not explicitly account for it in the measurement model.

\subsection{Scenario 2 }
\begin{figure}[tb]
    \centering
    \includegraphics[width=\columnwidth]{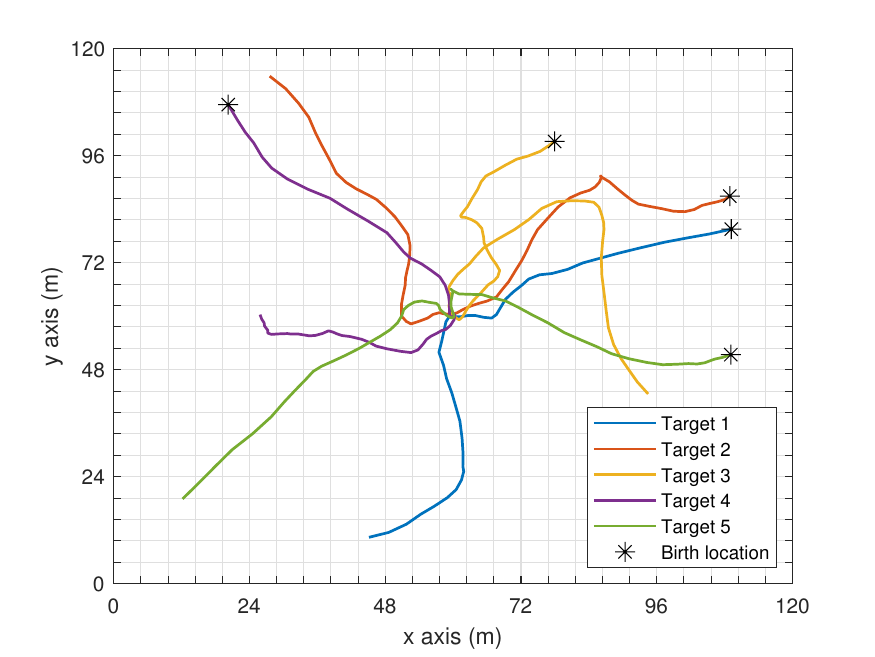}
    \caption{The simulated trajectories of the 5 different targets with their corresponding birth locations \cite{davies_information_2024}. The grid illustrates the resolution cells.}
    \label{fig:Trajectories_scen2}
\end{figure}

The second scenario is identical to the one in \cite{davies_information_2024} and the provided implementation was used in order to compare the proposed method with other state-of-the-art multi-target TBD filters. In this simulation, we consider a surveillance area of 120\,m \mathintext{\times} 120\,m divided into 25 \mathintext{\times} 25 equally spaced resolution cells. 5 targets start at different locations and at different times and move in a way such that they cross each other at the midpoint of the simulation at the centre of the surveillance region. This complex setting allows testing the multi-target tracking performance when multiple targets occupy the same resolution cell. The resulting trajectories of the simulated scenario 2 are shown in Fig. \ref{fig:Trajectories_scen2}.

\subsubsection{Simulation Setup}
\label{sec:Sim_scen2_setup}
In scenario 2, the same dynamical model (\ref{eq:target_dynamics}) as in Section \ref{sec:Sim_scen1_setup} was used with process noise variance \mathintext{q = 0.25\,\text{m}^2/\text{s}^3} and sample period \mathintext{T = 1}. In addition, the creation of measurements and the birth process were identical to \cite{davies_information_2024}. The influence of each target on the intensity of every resolution cell is given by the function\cite{davies_information_2024}:

\begin{equation}
    \makesubandsuper{h}{}{\pixelind}(\makesubandsuper{\vect{x}}{\timeind}{\targetind})= \frac{\phi}{(d^\pixelind(\makesubandsuper{\vect{x}}{\timeind}{\targetind}))^\beta + \epsilon} \;,
\end{equation}
where \mathintext{d^\pixelind(\makesubandsuper{\vect{x}}{\timeind}{\targetind})} is the distance from a target with state \mathintext{\makesubandsuper{\vect{x}}{\timeind}{\targetind}} to the centre of the \mathintext{\pixelind}-th resolution cell and \mathintext{\phi=400}, \mathintext{\epsilon=25}, and \mathintext{\beta=2}. \blue{This model considers a non-fluctuating amplitude, and at the target's location, the amplitude has a maximum value of 16.} The intensity in each resolution cell is found by the sum of all target contributions plus additional noise, i.e.:

\begin{equation}
    \makesubandsuper{z}{\timeind}{\pixelind} = \makesubandsuper{\sum}{\targetind=1}{M} \makesubandsuper{h}{}{\pixelind}(\makesubandsuper{\vect{x}}{\timeind}{\targetind}) + \makesubandsuper{w}{\timeind}{\pixelind} \;,
\end{equation}
where \mathintext{\makesubandsuper{w}{\timeind}{\pixelind}} is zero-mean Gaussian noise with variance \mathintext{\sigma_w = 1}. \blue{For the given noise and peak amplitude, using (50), this is equivalent to an $\blue{\mathrm{SNR} = 24.08\,\mathrm{dB}}$.} For the implementation of the IE-PHPMHT, the same densities \mathintext{p^m(y|\makesubandsuper{x}{\timeind}{\targetind})} and \mathintext{p^0(y)} as in scenario 1 were assumed. 

For the multi-Bernoulli birth, we assumed prior knowledge of the approximate birth locations, and targets were added with a birth probability of \mathintext{p_b = 1\times10^{-4}} and the states were initialised according to the Gaussian densities with means 
\mathintext{\makesubandsuper{\vect{x}}{b}{1} = \left[ 75\,\text{m} \ 0 \ 100\,\text{m} \ 0 \right]^\transp}, 
\mathintext{\makesubandsuper{\vect{x}}{b}{2} = \left[ 20\,\text{m} \ 0 \ 105\,\text{m} \ 0 \right]^\transp}, 
\mathintext{\makesubandsuper{\vect{x}}{b}{3} = \left[ 110\,\text{m} \ 0 \ 50\,\text{m} \ 0 \right]^\transp}, 
\mathintext{\makesubandsuper{\vect{x}}{b}{4} = \left[ 110\,\text{m} \ 0 \ 85\,\text{m} \ 0 \right]^\transp}, 
\mathintext{\makesubandsuper{\vect{x}}{b}{5} = \left[ 110\,\text{m} \ 0 \ 80\,\text{m} \ 0 \right]^\transp},
and covariances \mathintext{P_b = \operatorname{diag}(\left[10\,\mathrm{m}^2 \ 10\,\mathrm{m}^2\mathrm{s}^{-2} \ 10\,\mathrm{m}^2 \ 10\,\mathrm{m}^2\mathrm{s}^{-2}\right]^\transp)}. The probability of survival was set to \mathintext{p_s = 0.99}. Bernoulli components with a probability of existence \mathintext{>} 0.5 were considered detected, and those with a probability of existence \mathintext{<} 0.01 were removed.

A total of 81 time steps was simulated and the targets appeared at time steps 
\mathintext{\makesubandsuper{k}{b}{1} = 18}, 
\mathintext{\makesubandsuper{k}{b}{2} = 2}, 
\mathintext{\makesubandsuper{k}{b}{3} = 3}, 
\mathintext{\makesubandsuper{k}{b}{4} = 16}, 
\mathintext{\makesubandsuper{k}{b}{5} = 9}
and disappeared at
\mathintext{\makesubandsuper{k}{d}{1} = 64}, 
\mathintext{\makesubandsuper{k}{d}{2} = 79}, 
\mathintext{\makesubandsuper{k}{d}{3} = 78}, 
\mathintext{\makesubandsuper{k}{d}{4} = 77}, 
\mathintext{\makesubandsuper{k}{d}{5} = 71}.

We compared the results of the IE-PHPMHT filter with three of the information exchange TBD multi-Bernoulli filters proposed in \cite{davies_information_2024}. Those filters run a multi-Bernoulli filter for each resolution cell and exchange predicted mean and covariance for the update. Three versions with different linearisation methods were used for comparison. The information exchange multi-Bernoulli iterated posterior linearisation filter (IEMB-IPLF), the information exchange multi-Bernoulli unscented Kalman filter (IEMB-UKF), and the information exchange multi-Bernoulli extended Kalman filter (IEMB-EKF). The comparison also includes a UKF implementation of the multi-Bernoulli TBD filter in \cite{vo_joint_2010} which is referred to as (IMB-UKF) and the previously introduced TH-PHPMHT that uses SNR-based thresholding.

\subsubsection{Results}

\label{sec:Sim_scen2_results}
\begin{figure}[!t]
\centering
\subfloat[RMS GOSPA error]{\includegraphics[width=0.5\columnwidth]{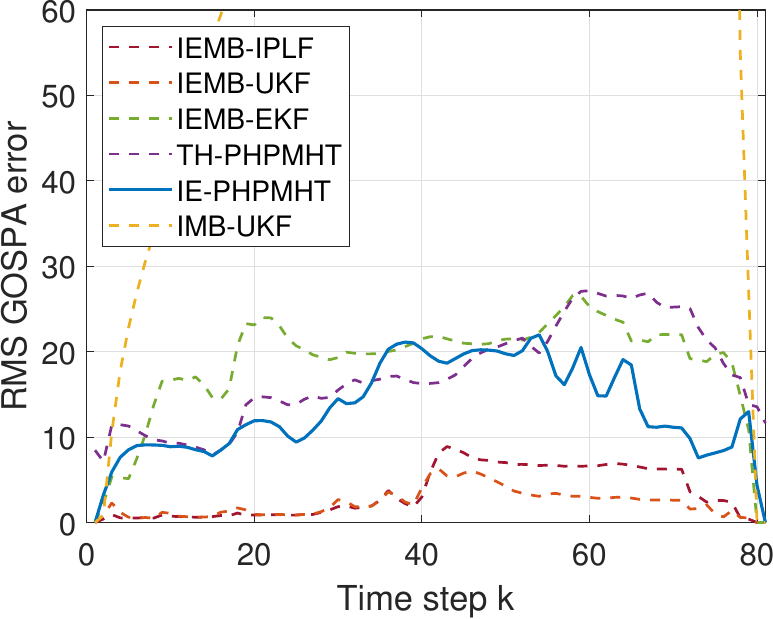}
\label{fig:GOSPA_scen2_a}}
\subfloat[RMS GOSPA localisation error]{\includegraphics[width=0.5\columnwidth]{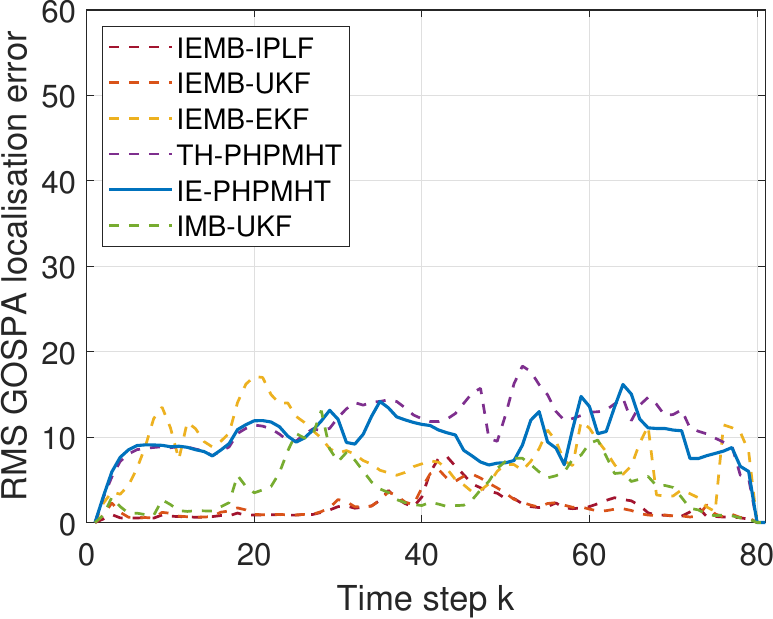}
\label{fig:GOSPA_scen2_b}}
\\
\subfloat[RMS GOSPA false target error]{\includegraphics[width=0.5\columnwidth]{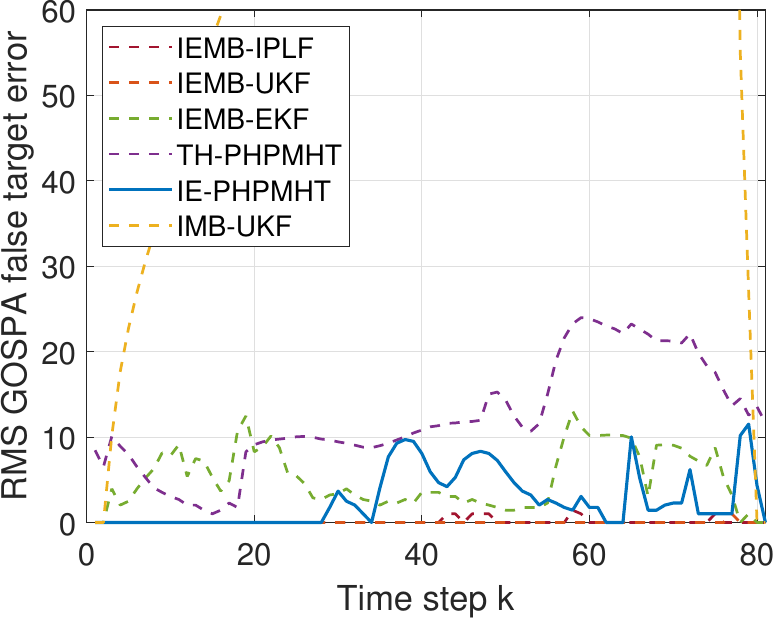}
\label{fig:GOSPA_scen2_c}}
\subfloat[RMS GOSPA missed target error]{\includegraphics[width=0.5\columnwidth]{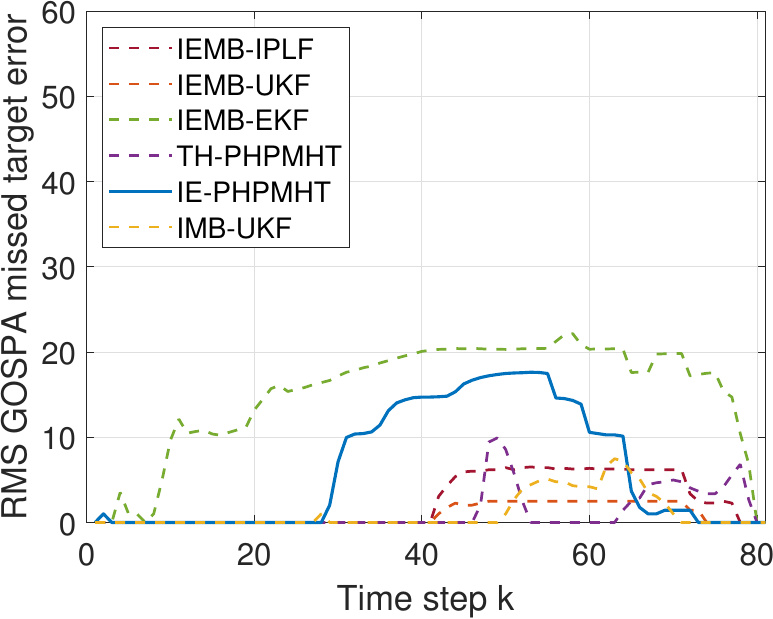}
\label{fig:GOSPA_scen2_d}}
\caption{Comparison of the RMS GOSPA errors and their decompositions for the different filters averaged over 100 Monte Carlo runs tested in scenario 2.}
\label{fig:GOSPA_scen2}
\end{figure}

Similarly to the previous Section \ref{sec:Sim_scen1_results}, the RMS GOSPA, plotted in Fig. \ref{fig:GOSPA_scen2}, was calculated for the performance assessment in scenario 2 with the only change being \mathintext{c = 3L} due to the altered dimensions of the simulation scenario. Recall, \mathintext{L} is the length of one resolution cell.

From Fig. \ref{fig:GOSPA_scen2_a}, we can see that the IEMB-UKF and IEMB-IPLF are the best performing filters, with the UKF implementation performing slightly better due to fewer missed targets. The IE-PHPMHT ranks third, although its performance is affected by higher localisation errors and an increased number of missed and false targets, particularly during the midpoint crossing in the simulation. The TH-PHPMHT and IEMB-EKF show higher errors than IE-PHPMHT, but these are still considerably lower than those of the IMB-UKF, which has the poorest performance due to a significant number of false targets dominating its total RMS GOSPA error.
Fig. \ref{fig:GOSPA_scen2_b} shows the localisation error, where IEMB-IPLF and IEMB-UKF exhibit considerably lower errors compared to the other filters. The IMB-UKF shows a low localisation error as well, but this is due to the high number of false targets. When more targets are retained within the filter, there is a higher chance that a target falls within the cut-off region, resulting in a lower localisation error. The IE-PHPMHT and TH-PHPMHT produce similar results, with only slight differences during the crossing scenario, as their methods differ in handling potential targets. In general, the two PHPMHT-based methods show the highest localisation error, likely due to the mismatch in their measurement models. The IEMB-EKF has higher localisation errors than the IEMB-IPLF and IEMB-UKF, likely caused by the use of the analytical linearisation.
In Fig. \ref{fig:GOSPA_scen2_c} we see the false target error, where IEMB-IPLF and IEMB-UKF perform best, maintaining very low false target errors throughout the simulation. The IE-PHPMHT performs similarly until approximately time step \mathintext{k=30}, when the targets come closer together, making it more complex for the filter to distinguish them. This leads to the creation of false targets due to merged measurements, likely caused by inaccurate estimates falling outside the cut-off region. Additionally, when targets disappear, the IE-PHPMHT shows an increased error, as it takes 1-2 time steps for the existence probability to drop and for the filter to remove a target. The IEMB-EKF produces more false targets than IE-PHPMHT, but fewer than TH-PHPMHT. The IMB-UKF performs the worst, as it consistently introduces new components at each time step without removing falsely initiated targets.
In Fig. \ref{fig:GOSPA_Scen1_SW1_d}, the missed target RMS GOSPA is shown, highlighting that IEMB-UKF misses the fewest targets, with some being missed after the crossing. The IEMB-IPLF shows similar behaviour but with a slightly higher number of false targets. The IE-PHPMHT again has difficulties distinguishing targets when their measurements intersect, leading to more missed targets. However, once the targets separate, the IE-PHPMHT successfully tracks them again, and the number of false targets decreases. The IEMB-EKF performs the worst, missing an increasing number of targets as the scenario becomes more complex, and this error persists throughout the simulation, resulting in the highest missed target error. The IMB-UKF and TH-PHPMHT show lower missed target errors, but this is largely due to the high number of false targets, increasing the likelihood that a false target is near the real target positions.

\begin{table}[tb]
\centering
\caption{Scenario 2 Average RMS GOSPA}
\label{tab:AVG_RMSGOSPA_Scen2}
\renewcommand{\arraystretch}{1.3}
\begin{tabular}{l||c|c|c|c}
\hline
\multicolumn{1}{l||}{\textbf{Method}} & \textbf{Error tot.} & \textbf{Error loc.} & \textbf{Error fal.} & \textbf{Error mis.} \\ \hline
IE-PHPMHT & 14.40 & 10.07 & 4.18 & 9.41 \\
TH-PHPMHT & 18.22 & 11.57 & 13.81 & 2.70 \\
IEMB-IPLF & 4.48 & 2.40 & 0.34 & 3.77 \\ 
IEMB-UKF & 2.76 & 2.33 & 0.11 & 1.48  \\ 
IEMB-EKF & 19.64 & 8.71 & 6.21 & 16.46 \\
IMB-UKF & 92.26 & 5.09 & 92.09 & 2.35 \\ \hline
\end{tabular}
\end{table}

The RMS GOSPA errors averaged over the entire simulation are summarised in Table \ref{tab:AVG_RMSGOSPA_Scen2}. On average, the total RMS GOSPA error of the IE-PHPMHT is lower than that of the TH-PHPMHT, IEMB-EKF, and IMB-UKF, though not as low as the IEMB-IPLF and IEMB-UKF. The latter two achieve notably fewer false targets and lower localisation errors, resulting in superior performance. Nonetheless, the IE-PHPMHT ranks third overall, indicating that despite the measurement model mismatch, it outperforms the IEMB-EKF, which relies on analytical linearisation. Additionally, the IE-PHPMHT handles target initiation and termination more effectively compared to the TH-PHPMHT, which uses the same tracking method but differs in track management. The TH-PHPMHT, aside from the IMB-UKF, has the highest number of false targets. This may be due to the low number of resolution cells used, leading to imprecise estimates of the Poisson rates and hence SNRs for the thresholding process. This causes worse performance compared to the IE-PHPMHT, which accounts for this uncertainty by incorporating the probability of existence and non-existence into the calculation. The IE-PHPMHT and TH-PHPMHT have the highest localisation errors due to a mismatch with the non-linear measurement model. However, this is only slightly higher than the IEMB-EKF, which uses analytical linearisation. In general, handling the non-linear nature of the measurements degrades performance, as the increased localisation error leads to more false and missed targets. Although the IMB-UKF uses sigma-point linearisation it has a very high number of false targets making it the worst performing filter in this scenario. This is likely due to the non-overlapping target assumption in the measurement model. 

\begin{table}[tb]
\centering
\caption{Comparison of the average runtimes in scenario 2}
\label{tab:AVG_runtime_Scen2}
\renewcommand{\arraystretch}{1.5}
\begin{tabular}{c||c}
\hline
\textbf{Method} & \textbf{Average runtime per time step (s)} \\ \hline
IE-PHPMHT & \underline{0.03} \\ 
TH-PHPMHT & 0.12 \\ 
IEMB-IPLF & 1.11 \\ 
IEMB-UKF & 0.44  \\ 
IEMB-EKF & 0.21  \\ 
IMB-UKF & 2.97  \\ \hline
\end{tabular}
\end{table}

Lastly, we evaluated the computational costs of the different methods used in scenario 2. The average runtime per time step across the Monte Carlo runs is presented in Table \ref{tab:AVG_runtime_Scen2}. The fastest execution time was achieved by the IE-PHPMHT, followed by the TH-PHPMHT. The longer runtime of the TH-PHPMHT, compared to the IE-PHPMHT, is due to the higher number of false targets, requiring state estimations for a larger number of potential targets. The performance difference within the three IEMB filters coincides with the results reported in \cite{davies_information_2024}. The IEMB-IPLF is the slowest as it performs iterated statistical linear regression, followed by the IEMB-UKF which uses iterations over the sigma-points for the linearisation, and lastly the IEMB-EKF with the analytical linearisation. Note that the maximum number of iterations for the IPLF implementation was set to 5 to get a good trade-off between performance and computational effort. The slowest method is the IMB-UKF which is likely caused by the high number of false targets that need to be processed. Overall, this underlines the low computational burden of the PHPMHT based methods. Moreover, it should be noted that the difference in the computational burden becomes more significant with an increasing number of resolution cells as the PHPMHT based methods scale only linearly with the number of resolution cells.
\section{Conclusion}
\label{sec:Conclusion}
In this paper, we have introduced the integrated existence Poisson histogram-probabilistic multi-hypothesis tracker for TBD of multiple targets. For the update step, we incorporated a new model for the generation of measurements that integrates target existence probabilities, overcoming the limitation of assuming a fixed and known number of targets. It has been derived by introducing auxiliary variables to the PPP measurement model. Additionally, we have proposed a mixture merging approximation of the Poisson measurement rate distribution by minimising the KLD to retain a closed-form solution.

In conclusion, the integrated track management of the IE-PHPMHT demonstrated improved performance compared to previous work (TH-PHPMHT) which is based on additional track management, notably in reducing false and missed targets and handling target amplitude fluctuations. Compared to other state-of-the-art multi-target tracking methods, the IE-PHPMHT achieved competitive results in terms of the average RMS GOSPA, with significantly reduced computational costs. This makes it particularly suitable in scenarios with a high number of resolution cells and targets. Overall, the IE-PHPMHT balances computational complexity and tracking accuracy, showing improvements in direct numerical comparisons within the PHPMHT framework.

\blue{Future work includes the time-dependent modelling of target returns to track the average target signal strength. Moreover, it is relevant to carry out further research exploring different, systematic target birth and initialisation strategies. Another line of future work is the application and evaluation of the algorithm using real-world data.}

\blue{\appendix[Proof of Lemma 2]}
\everymath{\color{black}}
\allowdisplaybreaks
\blue{In this Appendix, we provide the proof of Lemma 2. For the EM auxiliary function in \eqref{eq:aux} we need to derive the conditional expectation of the logarithm of the complete likelihood over the missing data. Starting from \eqref{eq:aux_comp_unsolved}, and following \eqref{eq:aux_div_general}, we divide the function into the components that depend solely on the states \mathintext{X_\timeind} and on the Poisson rates \mathintext{\Lambda_\timeind}:}
\small
\begin{align}
\label{eq:app_aux_X_unsolved}
    \makesubandsuper{\aux}{\timeind,X}{\itind} & \left(X_\timeind,\Lambda_\timeind|\hat{X}_\timeind^{(\itind-1)},\hat{\Lambda}_\timeind^{(\itind-1)}\right) \nonumber\\
    =& \makesubandsuper{\sum}{\targetind=1}{\Targetind} \log \left\{ p(\makesubandsuper{\vect{x}}{\timeind}{\targetind}) \right\} 
    + \int \left[ \makesubandsuper{\sum}{\pixelindx=1}{\Pixelindx} \makesubandsuper{\sum}{r=1}{\makesubandsuper{n}{\timeind}{\pixelind}}
    \log \left\{ p^{i,\makesubandsuper{m}{\timeind}{\pixelind,r}}\left( \makesubandsuper{\vect{y}}{\timeind}{\pixelind,r}|\makesubandsuper{\vect{x}}{\timeind}{\makesubandsuper{m}{\timeind}{\pixelind,r}}\right) \right\} \right] \nonumber \\
    &\times \makesubandsuper{\prod}{\pixelindx=1}{\Pixelindx} n_\timeind^{\pixelind}! \prod_{r=1}^{n_\timeind^{\pixelind}} \frac{ \, \makesubandsuper{\hat{\lambda}}{\timeind}{\makesubandsuper{m}{\timeind}{\pixelind,r},{(\itind-1)}} \makesubandsuper{p}{}{i,\makesubandsuper{m}{\timeind}{\pixelind,r}}\left( \makesubandsuper{\vect{y}}{\timeind}{\pixelind,r} | \makesubandsuper{\hat{\vect{x}}}{\timeind}{\makesubandsuper{m}{\timeind}{\pixelind,r},{(\itind-1)}} \right)}{\blue{\makesubandsuper{\hat{\Bar{\nu}}}{\timeind}{\pixelind,{(\itind-1)}}}}
    \ \delta \mathbb{Y}_\timeind \;,
\end{align}
\normalsize
\small
\begin{align}
\label{eq:app_aux_Lambda_unsolved}
    \makesubandsuper{\aux}{\timeind,\Lambda}{\itind} & \left(X_\timeind,\Lambda_\timeind|\hat{X}_\timeind^{(\itind-1)},\hat{\Lambda}_\timeind^{(\itind-1)}\right) \nonumber\\
    =& \makesubandsuper{\sum}{\targetind=0}{\Targetind} 
    \Big[ \log \left\{ p(\makesubandsuper{\lambda}{\timeind}{\targetind} )\right\} - \makesubandsuper{\lambda}{\timeind}{\targetind} \Big]
    + \int \left[ \makesubandsuper{\sum}{\pixelindx=1}{\Pixelindx} \makesubandsuper{\sum}{r=1}{\makesubandsuper{n}{\timeind}{\pixelind}} \log \left\{ \makesubandsuper{\lambda}{\timeind}{\makesubandsuper{m}{\timeind}{\pixelind,r}} \right\} \right] \nonumber \\
    &\times \makesubandsuper{\prod}{\pixelindx=1}{\Pixelindx} n_\timeind^{\pixelind}! \prod_{r=1}^{n_\timeind^{\pixelind}} \frac{ \, \makesubandsuper{\hat{\lambda}}{\timeind}{\makesubandsuper{m}{\timeind}{\pixelind,r},{(\itind-1)}} \makesubandsuper{p}{}{i,\makesubandsuper{m}{\timeind}{\pixelind,r}}\left( \makesubandsuper{\vect{y}}{\timeind}{\pixelind,r} | \makesubandsuper{\hat{\vect{x}}}{\timeind}{\makesubandsuper{m}{\timeind}{\pixelind,r},{(\itind-1)}} \right)}{\blue{\makesubandsuper{\hat{\Bar{\nu}}}{\timeind}{\pixelind,{(\itind-1)}}}}
    \ \delta \mathbb{Y}_\timeind \;.
\end{align}
\normalsize
\blue{\subsection{Derivation of the Poisson rate EM auxiliary function}}
\blue{First, we derive the part of the auxiliary function that depends on the Poisson rates. Therefore, we have to calculate the integral in \eqref{eq:app_aux_Lambda_unsolved}, which we denote here as \mathintext{\mathcal{I}_\Lambda} for brevity. We insert the definition of a sequence of set integrals \cite{mahler_advances_2014} over a sequence of sets:}
\small
\begin{align}
\mathcal{I}_\Lambda & =\sum_{n_{k}^{1}=0}^{\infty}...\sum_{n_{k}^{I}=0}^{\infty}\left[\prod_{i=1}^{I}\frac{1}{n_{k}^{i}!}\right]\sum_{m_{k}^{1,1}=0}^{M_{k}}....\sum_{m_{k}^{I,n_{k}^{i}}=0}^{M_{k}} \nonumber \\
&\int...\int
 \left[\sum_{i=1}^{I}\sum_{r=1}^{n_{k}^{i}}\log\lambda_{k}^{m_{k}^{i,r}}\right] \nonumber\\
 & \times\prod_{i=1}^{I}n_{k}^{i}!\prod_{r=1}^{n_{k}^{i}}\frac{\hat{\lambda}_{k}^{m_{k}^{i,r},(p-1)}p^{i,m_{k}^{i,r}}\left(y_{k}^{i,r}|\hat{x}^{m_{k}^{i,r},(p-1)}\right)}{\hat{\Bar{\nu}}_k^{i,(p-1)}} \nonumber\\
 & \times dy_{k}^{1,1}...dy_{k}^{I,n_{k}^{I}} \;.
\end{align}
\normalsize
\blue{We observe that the factorials cancel out. Note that the multi-object densities in \eqref{eq:p_miss_complete} are for a sequence of sets with their cardinality fixed, each with \mathintext{\makesubandsuper{n}{\timeind}{\pixelind}}. Otherwise, the multi-object density is zero. Hence, for other cardinalities, the terms in the set integrals are zero. This yields the following expression:}
\small
\begin{align}
\mathcal{I}_\Lambda & =\sum_{m_{k}^{1,1}=0}^{M_{k}}....\sum_{m_{k}^{I,n_{k}^{i}}=0}^{M_{k}}\int...\int
\times\left[\sum_{i=1}^{I}\sum_{r=1}^{n_{k}^{i}}\log\lambda_{k}^{m_{k}^{i,r}}\right] \nonumber\\
 & \times\prod_{i=1}^{I}\prod_{r=1}^{n_{k}^{i}}\frac{\hat{\lambda}_{k}^{m_{k}^{i,r},(p-1)}p^{i,m_{k}^{i,r}}\left(y_{k}^{i,r}|\hat{x}^{m_{k}^{i,r},(p-1)}\right)}{\hat{\Bar{\nu}}_k^{i,(p-1)}} dy_{k}^{1,1}...dy_{k}^{I,n_{k}^{I}} \;.
\end{align}
\normalsize
\blue{Next, we simplify the integrals:}
\small
\begin{align}
\mathcal{I}_\Lambda & =\sum_{m_{k}^{1,1}=0}^{M_{k}}....\sum_{m_{k}^{I,n_{k}^{i}}=0}^{M_{k}}
 \left[\sum_{i=1}^{I}\sum_{r=1}^{n_{k}^{i}}\log\lambda_{k}^{m_{k}^{i,r}}\right] \nonumber\\
 & \times\prod_{i=1}^{I}\prod_{r=1}^{n_{k}^{i}}\frac{\hat{\lambda}_{k}^{m_{k}^{i,r},(p-1)}\int_{A^i} p^{i,m_{k}^{i,r}}\left(y|\hat{x}^{m_{k}^{i,r},(p-1)}\right)dy}{\hat{\Bar{\nu}}_k^{i,(p-1)}} \;.
\end{align}
\normalsize
\blue{We change the indices of the multiplications to take the next steps:}
\small
\begin{align}
\mathcal{I}_\Lambda & =\sum_{m_{k}^{1,1}=0}^{M_{k}}....\sum_{m_{k}^{I,n_{k}^{i}}=0}^{M_{k}}
\left[\sum_{i=1}^{I}\sum_{r=1}^{n_{k}^{i}}\log\lambda_{k}^{m_{k}^{i,r}}\right] \nonumber\\
 & \times\prod_{\Tilde{i}=1}^{I}\prod_{\Tilde{r}=1}^{n_{k}^{i}}\frac{\hat{\lambda}_{k}^{m_{k}^{\Tilde{i},\Tilde{r}},(p-1)}\int_{A^i} p^{\Tilde{i},m_{k}^{\Tilde{i},\Tilde{r}}}\left(y|\hat{x}^{m_{k}^{\Tilde{i},\Tilde{r}},(p-1)}\right)dy}{\hat{\Bar{\nu}}_k^{\Tilde{i},(p-1)}} \;.
\end{align}
\normalsize
\blue{Then, we reorganize the multiplications such that we first multiply the term with index $i$,$r$ and then we multiply over the rest of the indices:}
\small
\begin{align}
\mathcal{I}_\Lambda & =\sum_{m_{k}^{1,1}=0}^{M_{k}}....\sum_{m_{k}^{I,n_{k}^{i}}=0}^{M_{k}} \sum_{i=1}^{I}\sum_{r=1}^{n_{k}^{i}} \nonumber\\
& \times \left[\frac{\hat{\lambda}_{k}^{m_{k}^{i,r},(p-1)}\int_{A^i} p^{i,m_{k}^{i,r}}\left(y|\hat{x}^{m_{k}^{i,r},(p-1)}\right)dy}{\hat{\Bar{\nu}}_k^{i,(p-1)}}\log\lambda_{k}^{m_{k}^{i,r}}\right.\nonumber\\
 & \times \left.\prod_{\substack{\Tilde{i}=1 \\ \Tilde{i}\neq i}}^{I}\prod_{\substack{\Tilde{r}=1 \\ \Tilde{r}\neq r}}^{n_{k}^{i}}\frac{\hat{\lambda}_{k}^{m_{k}^{\Tilde{i},\Tilde{r}},(p-1)}\int_{A^i} p^{\Tilde{i},m_{k}^{\Tilde{i},\Tilde{r}}}\left(y|\hat{x}^{m_{k}^{\Tilde{i},\Tilde{r}},(p-1)}\right)dy}{\hat{\Bar{\nu}}_k^{\Tilde{i},(p-1)}}\right] \;.
\end{align}
\normalsize
\blue{Changing the order of the summation yields:}
\small
\begin{align}
\mathcal{I}_\Lambda & =\sum_{i=1}^{I}\sum_{r=1}^{n_{k}^{i}} \nonumber\\
& \times \left[\sum_{m_{k}^{i,r}=0}^{M_{k}}\frac{\hat{\lambda}_{k}^{m_{k}^{i,r},(p-1)}\int_{A^i} p^{i,m_{k}^{i,r}}\left(y|\hat{x}^{m_{k}^{i,r},(p-1)}\right)dy}{\hat{\Bar{\nu}}_k^{i,(p-1)}}\log\lambda_{k}^{m_{k}^{i,r}}\right. \nonumber\\
 & \left.\times\prod_{\substack{\Tilde{i}=1 \\ \Tilde{i}\neq i}}^{I}\prod_{\substack{\Tilde{r}=1 \\ \Tilde{r}\neq r}}^{n_{k}^{i}}\sum_{m_{k}^{\Tilde{i},\Tilde{r}}=0}^{M_{k}}\frac{\hat{\lambda}_{k}^{m_{k}^{\Tilde{i},\Tilde{r}},(p-1)}\int_{A^i} p^{\Tilde{i},m_{k}^{\Tilde{i},\Tilde{r}}}\left(y|\hat{x}^{m_{k}^{\Tilde{i},\Tilde{r}},(p-1)}\right)dy}{\hat{\Bar{\nu}}_k^{\Tilde{i},(p-1)}}\right] \;.
\end{align}
\normalsize
\blue{Now, using \eqref{eq:nu_bar_hat}, the term in the last line above is equal to one which results in:}
\small
\begin{align}
\mathcal{I}_\Lambda & =\sum_{i=1}^{I}\sum_{r=1}^{n_{k}^{i}}\sum_{m_{k}^{i,r}=0}^{M_{k}} \nonumber\\
& \times \frac{\hat{\lambda}_{k}^{m_{k}^{i,r},(p-1)}\int_{A^i} p^{i,m_{k}^{i,r}}\left(y|\hat{x}^{m_{k}^{i,r},(p-1)}\right)dy}{\hat{\Bar{\nu}}_k^{i,(p-1)}}\log\lambda_{k}^{m_{k}^{i,r}}\\
 & =\sum_{i=1}^{I}\sum_{r=1}^{n_{k}^{i}}\frac{\sum_{m=0}^{M_{k}}\log\left\{\lambda_{k}^{m}\right\}\hat{\lambda}_{k}^{m,(p-1)}\int_{A^i} p^{i,m}\left(y|\hat{x}^{m,(p-1)}\right)dy}{\hat{\Bar{\nu}}_k^{i,(p-1)}}\\
 & =\sum_{m=0}^{M_{k}}\log\left\{\lambda_{k}^{m}\right\}\hat{\lambda}_{k}^{m,(p-1)}\sum_{i=1}^{I}\sum_{r=1}^{n_{k}^{i}}\frac{\int_{A^i} p^{i,m}\left(y|\hat{x}^{m,(p-1)}\right)dy}{\hat{\Bar{\nu}}_k^{i,(p-1)}}\\
 & = \label{eq:I_Lambda_final}\sum_{m=0}^{M_{k}}\log\left\{\lambda_{k}^{m}\right\}\hat{\lambda}_{k}^{m,(p-1)}\sum_{i=1}^{I}n_{k}^{i}\frac{\int_{A^i} p^{i,m}\left(y|\hat{x}^{m,(p-1)}\right)dy}{\hat{\Bar{\nu}}_k^{i,(p-1)}} \;.
\end{align}
\normalsize
\blue{Inserting the solution to the sequence of set integrals \mathintext{\mathcal{I}_\Lambda}, given in \eqref{eq:I_Lambda_final}, back into \eqref{eq:app_aux_Lambda_unsolved}, we obtain the final result for the auxiliary function that depends on the Poisson rates:}
\small
\begin{align}
    \makesubandsuper{\aux}{\timeind,\Lambda}{\itind} & \left(X_\timeind,\Lambda_\timeind|\hat{X}_\timeind^{(\itind-1)},\hat{\Lambda}_\timeind^{(\itind-1)}\right) \nonumber\\
    =& \makesubandsuper{\sum}{\targetind=0}{\Targetind} \Biggl[ \log \left\{ p(\makesubandsuper{\lambda}{\timeind}{\targetind}) \right\} -\makesubandsuper{\lambda}{\timeind}{\targetind} + \log \left\{ \makesubandsuper{\lambda}{\timeind}{\targetind} \right\} \nonumber \\
    &\times \makesubandsuper{\sum}{\pixelind=1}{\Pixelindx} \makesubandsuper{n}{\timeind}{\pixelind}
    \frac{\makesubandsuper{\hat{\lambda}}{\timeind}{\targetind,(\itind-1)}  \int_{A^\pixelind} p^{i,\targetind}\left( y|\makesubandsuper{\hat{\vect{x}}}{\timeind}{\targetind,(\itind-1)} \right) dy}{\makesubandsuper{\hat{\bar{\nu}}}{\timeind}{\pixelind,(\itind-1)}} \Biggr] \\
    =& \makesubandsuper{\sum}{\targetind=0}{\Targetind} \left[ \log \left\{ p(\makesubandsuper{\lambda}{\timeind}{\targetind}) \right\} 
    + \log \left\{ \text{e}^{-\makesubandsuper{\lambda}{\timeind}{\targetind}} {\left(\makesubandsuper{\lambda}{\timeind}{\targetind}\right)}^{\makesubandsuper{\Bar{n}}{\timeind}{\targetind}} \right\} \right] \;,
\end{align}
\normalsize
\blue{with}
\small
\begin{align}
    \makesubandsuper{\Bar{n}}{\timeind}{\targetind} = \makesubandsuper{\sum}{\pixelind=1}{\Pixelind} \makesubandsuper{n}{\timeind}{\pixelind}
    \frac{\makesubandsuper{\hat{\lambda}}{\timeind}{\targetind,(\itind-1)}  \int_{A^\pixelind} p^{i,\targetind}\left( y|\makesubandsuper{\hat{\vect{x}}}{\timeind}{\targetind,(\itind-1)} \right) dy}{\makesubandsuper{\sum}{s=0}{\Targetind} \makesubandsuper{\hat{\lambda}}{\timeind}{s,(\itind-1)} \int_{A^\pixelind} p^{i,s}\left( y|\makesubandsuper{\hat{\vect{x}}}{\timeind}{s,(\itind-1)} \right) dy} \;.
\end{align}
\normalsize
\blue{\subsection{Derivation of the state EM auxiliary function}}
\blue{In the same manner, we proceed with the calculation of the sequence of set integrals of the state-dependent auxiliary function \eqref{eq:app_aux_X_unsolved}. 
We insert the definition of the sequence of set integrals \cite{mahler_advances_2014} over a sequence of sets:}
\small
\begin{align}
\mathcal{I}_X & =\sum_{n_{k}^{1}=0}^{\infty}...\sum_{n_{k}^{I}=0}^{\infty}\left[\prod_{i=1}^{I}\frac{1}{n_{k}^{i}!}\right]\sum_{m_{k}^{1,1}=0}^{M_{k}}....\sum_{m_{k}^{I,n_{k}^{i}}=0}^{M_{k}} \nonumber \\
&\int...\int
 \left[\sum_{i=1}^{I}\sum_{r=1}^{n_{k}^{i}}\log p^{i,m_k^{i,r}} \left( y_k^{i,r} | x_k^{m_k^{i,r}} \right) \right] \nonumber\\
 & \times\prod_{i=1}^{I}n_{k}^{i}!\prod_{r=1}^{n_{k}^{i}}\frac{\hat{\lambda}_{k}^{m_{k}^{i,r},(p-1)}p^{i,m_{k}^{i,r}}\left(y_{k}^{i,r}|\hat{x}^{m_{k}^{i,r},(p-1)}\right)}{\hat{\Bar{\nu}}_k^{i,(p-1)}} \nonumber\\
 & \times dy_{k}^{1,1}...dy_{k}^{I,n_{k}^{I}} \;.
\end{align}
\normalsize
\blue{Again, we note that the factorials cancel out, and we consider a fixed cardinality. Using the same reasoning as before, we obtain:}
\small
\begin{align}
\mathcal{I}_X &= \sum_{m_{k}^{1,1}=0}^{M_{k}}....\sum_{m_{k}^{I,n_{k}^{i}}=0}^{M_{k}}\int...\int
\left[\sum_{i=1}^{I}\sum_{r=1}^{n_{k}^{i}}\log p^{i,m_k^{i,r}} \left( y_k^{i,r} | x_k^{m_k^{i,r}} \right) \right] \nonumber\\
& \times\prod_{i=1}^{I}\prod_{r=1}^{n_{k}^{i}}\frac{\hat{\lambda}_{k}^{m_{k}^{i,r},(p-1)}p^{i,m_k^{i,r}}\left(y_{k}^{i,r}|\hat{x}^{m_k^{i,r},(p-1)}\right)}{\hat{\Bar{\nu}}_k^{i,(p-1)}} dy_{k}^{1,1}...dy_{k}^{I,n_{k}^{I}} \;.
\end{align}
\normalsize
\blue{Now, we first multiply the term with index $i$,$r$ and then we multiply over the rest of the indices:}
\small
\begin{align}
\mathcal{I}_X &= \sum_{m_{k}^{1,1}=0}^{M_{k}}....\sum_{m_{k}^{I,n_{k}^{i}}=0}^{M_{k}}\int...\int
\left[\sum_{i=1}^{I}\sum_{r=1}^{n_{k}^{i}}\log \left\{ p^{i,m_k^{i,r}} \left( y_k^{i,r} | x_k^{m_k^{i,r}} \right) \right\} \right. \nonumber\\
& \times \frac{\hat{\lambda}_{k}^{m_{k}^{i,r},(p-1)}p^{i,m_k^{i,r}}\left(y_{k}^{i,r}|\hat{x}^{m_k^{i,r},(p-1)}\right)}{\hat{\Bar{\nu}}_k^{i,(p-1)}} \nonumber \\
& \left.\times\prod_{\substack{\Tilde{i}=1 \\ \Tilde{i}\neq i}}^{I}\prod_{\substack{\Tilde{r}=1 \\ \Tilde{r}\neq r}}^{n_{k}^{i}}\frac{\hat{\lambda}_{k}^{m_{k}^{\Tilde{i},\Tilde{r}},(p-1)}p^{\Tilde{i},m_k^{\Tilde{i},\Tilde{r}}}\left(y_{k}^{\Tilde{i},\Tilde{r}}|\hat{x}^{m_k^{\Tilde{i},\Tilde{r}},(p-1)}\right)}{\hat{\Bar{\nu}}_k^{\Tilde{i},(p-1)}} \right] dy_{k}^{1,1}...dy_{k}^{I,n_{k}^{I}} \;.
\end{align}
\normalsize
\blue{Changing the order of the summation yields:}
\small
\begin{align}
\mathcal{I}_X &= \int...\int
\left[\sum_{i=1}^{I}\sum_{r=1}^{n_{k}^{i}} \left[ \sum_{m_{k}^{i,r}=0}^{M_{k}} \log \left\{ p^{i,m_k^{i,r}} \left( y_k^{i,r} | x_k^{m_k^{i,r}} \right) \right\} \right. \right. \nonumber\\
& \times \frac{\hat{\lambda}_{k}^{m_{k}^{i,r},(p-1)}p^{i,m_k^{i,r}}\left(y_{k}^{i,r}|\hat{x}^{m_k^{i,r},(p-1)}\right)}{\hat{\Bar{\nu}}_k^{i,(p-1)}} \nonumber \\
& \left.\left.\times\prod_{\substack{\Tilde{i}=1 \\ \Tilde{i}\neq i}}^{I}\prod_{\substack{\Tilde{r}=1 \nonumber\\ \Tilde{r}\neq r}}^{n_{k}^{i}} \sum_{m_{k}^{\Tilde{i},\Tilde{r}}=0}^{M_{k}} \frac{\hat{\lambda}_{k}^{m_{k}^{\Tilde{i},\Tilde{r}},(p-1)}p^{\Tilde{i},m_k^{\Tilde{i},\Tilde{r}}}\left(y_{k}^{\Tilde{i},\Tilde{r}}|\hat{x}^{m_k^{\Tilde{i},\Tilde{r}},(p-1)}\right)}{\hat{\Bar{\nu}}_k^{\Tilde{i},(p-1)}} \right] \right] \\
& \times \ dy_{k}^{1,1}...dy_{k}^{I,n_{k}^{I}} \;.
\end{align}
\normalsize
\blue{Using the identity of \mathintext{\hat{\Bar{\nu}}_k^{i,(p-1)}} as before, the last line above is equal to one which further simplifies to:}
\small
\begin{align}
\mathcal{I}_X &= \int...\int
\left[\sum_{i=1}^{I}\sum_{r=1}^{n_{k}^{i}}  \sum_{m_{k}^{i,r}=0}^{M_{k}} \log \left\{ p^{i,m_k^{i,r}} \left( y_k^{i,r} | x_k^{m_k^{i,r}} \right) \right\} \right.  \nonumber\\
& \left. \times \frac{\hat{\lambda}_{k}^{m_{k}^{i,r},(p-1)}p^{i,m_k^{i,r}}\left(y_{k}^{i,r}|\hat{x}^{m_k^{i,r},(p-1)}\right)}{\hat{\Bar{\nu}}_k^{i,(p-1)}} \right] dy_{k}^{1,1}...dy_{k}^{I,n_{k}^{I}} \;.
\end{align}
\normalsize
\blue{Next, we simplify the integrals and rearrange the terms:}
\small
\begin{align}
\mathcal{I}_X &= \sum_{i=1}^{I}\sum_{r=1}^{n_{k}^{i}}  \sum_{m_{k}^{i,r}=0}^{M_{k}} \frac{\hat{\lambda}_{k}^{m_{k}^{i,r},(p-1)}}{\hat{\Bar{\nu}}_k^{i,(p-1)}} \int_{A^i} \log \left\{ p^{i,m_k^{i,r}} \left( y | x_k^{m_k^{i,r}} \right) \right\}  \nonumber\\
& \times p^{i,m_k^{i,r}}\left(y|\hat{x}^{m_k^{i,r},(p-1)}\right) dy \\
&= \sum_{i=1}^{I}\sum_{r=1}^{n_{k}^{i}}  \sum_{m=0}^{M_{k}} \frac{\hat{\lambda}_{k}^{m,(p-1)}}{\hat{\Bar{\nu}}_k^{i,(p-1)}} \int_{A^i} \log \left\{ p^{i,m} \left( y | x_k^{m} \right) \right\}  \nonumber\\
& \times p^{i,m}\left(y|\hat{x}^{m,(p-1)}\right) dy \\
&= \label{eq:I_X_final} \sum_{m=0}^{M_{k}} \left[ \hat{\lambda}_{k}^{m,(p-1)} \sum_{i=1}^{I} \frac{n_k^i}{\hat{\Bar{\nu}}_k^{i,(p-1)}} \int_{A^i} \log \left\{ p^{i,m} \left( y | x_k^{m} \right) \right\} \right.  \nonumber\\
& \left.\times p^{i,m}\left(y|\hat{x}^{m,(p-1)}\right) dy \right] \;.
\end{align}
\normalsize
\blue{Inserting the solution to the sequence of set integrals \mathintext{\mathcal{I}_X}, given in \eqref{eq:I_X_final}, back into \eqref{eq:app_aux_X_unsolved}, we get the final result for the state-dependent auxiliary function:}
\small
\begin{align}
    \makesubandsuper{\aux}{\timeind,X}{\itind} & \left(X_\timeind,\Lambda_\timeind|\hat{X}_\timeind^{(\itind-1)},\hat{\Lambda}_\timeind^{(\itind-1)}\right) \nonumber\\
    =& \makesubandsuper{\sum}{\targetind=0}{\Targetind} \Biggl[ \log \left\{ p(\makesubandsuper{\vect{x}}{\timeind}{\targetind}) \right\} + \makesubandsuper{\hat{\lambda}}{\timeind}{\targetind,(\itind-1)}
    \makesubandsuper{\sum}{\pixelind=1}{\Pixelindx} \frac{\makesubandsuper{n}{\timeind}{\pixelind}}{\makesubandsuper{\hat{\Bar{\nu}}}{\timeind}{\pixelind,(\itind-1)}} \nonumber \\
    & \times \int_{A^\pixelind} \log \left\{ p^{i,\targetind} \left( y|\makesubandsuper{\vect{x}}{\timeind}{\targetind} \right) \right\} p^{i,\targetind} \left( y|\makesubandsuper{\hat{\vect{x}}}{\timeind}{\targetind,(\itind-1)} \right) d y \Biggr] \;.
\end{align}
\normalsize
\blue{This finishes the proof of Lemma 2.}

\appendices
\ifCLASSOPTIONcaptionsoff
  \newpage
\fi



\end{document}